\documentclass[11pt]{article}
\usepackage[utf8]{inputenc}
\usepackage{mathtools}
\usepackage{hyperref}
\usepackage{dsfont}
\usepackage{amsmath,amssymb}
\usepackage[superscript,biblabel]{cite}
\usepackage{authblk}
\usepackage[margin=1.3in]{geometry}
\usepackage[labelfont=bf]{caption}
\usepackage{longtable}
\usepackage{tabu}
\usepackage{colortbl}
\usepackage{listings}
\usepackage{booktabs}
\usepackage{xcolor}
\PassOptionsToPackage{hyphens}{url}\usepackage{hyperref}

\begin{document}

\title{
Methods for Population Adjustment with Limited Access to Individual Patient Data: A Review and Simulation Study}
\author[1,2]{Antonio Remiro-Azócar}
\author[1,3,4]{Anna Heath}
\author[1]{Gianluca Baio}
\affil[1]{Department of Statistical Science, University College London}
\affil[2]{Quantitative Research, Statistical Outcomes Research \& Analytics (SORA)}
\affil[3]{Child Health Evaluative Sciences, The Hospital for Sick Children}
\affil[4]{Dalla Lana School of Public Health, University of Toronto}
\date{v0.8, \today}

\maketitle

\begin{abstract}
Population-adjusted indirect comparisons estimate treatment effects when access to individual patient data is limited and there are cross-trial differences in effect modifiers. Health technology assessment agencies are accepting evaluations that use these methods across a diverse range of therapeutic areas. Popular methods include matching-adjusted indirect comparison (MAIC) and simulated treatment comparison (STC). There is limited formal evaluation of these methods and whether they can be used to accurately compare treatments. Thus, we undertake a comprehensive simulation study to compare standard unadjusted indirect comparisons, MAIC and STC across 162 scenarios. This simulation study assumes that the trials are investigating survival outcomes and measure continuous covariates, with the log hazard ratio as the measure of effect --- one of the most widely used setups in health technology assessment applications. The simulation scenarios vary the trial sample size, prognostic variable effects, interaction effects, covariate correlations and covariate overlap. MAIC yields unbiased treatment effect estimates under no failures of assumptions. The typical usage of STC produces bias because it targets a conditional treatment effect where the target estimand should be a marginal treatment effect. The incompatibility of estimates in the indirect comparison leads to bias as the measure of effect is non-collapsible. Standard indirect comparisons are systematically biased, particularly under stronger covariate imbalance and interaction effects. Standard errors and coverage rates are often valid in MAIC but the robust sandwich variance estimator underestimates variability where effective sample sizes are small. Interval estimates for the standard indirect comparison are too narrow and STC suffers from bias-induced undercoverage. MAIC provides the most accurate estimates and, with lower degrees of covariate overlap, its bias reduction outweighs the loss in effective sample size and precision under no failures of assumptions. An important future objective is the development of an alternative formulation to STC that targets a marginal treatment effect.
\end{abstract}

\pagebreak

\subsection*{Highlights}

\paragraph{What is already known?}

\begin{itemize}
\item Population adjustment methods such as matching-adjusted indirect comparison \\(MAIC) and simulated treatment comparison (STC) are increasingly used to compare treatments in health technology assessments.

\item Such methods estimate treatment effects when there are differences in effect modifiers across trials and when access to patient-level data is limited.   
\end{itemize}

\paragraph{What is new?}

\begin{itemize}
\item We present a comprehensive simulation study which benchmarks the performance of MAIC and STC against the standard unadjusted comparison across 162 scenarios.  

\item The simulation study provides the most extensive evaluation of population adjustment methods to date and informs the circumstances under which the methods should be applied. 
\end{itemize}

\paragraph{Potential impact for readers}

\begin{itemize}
\item In the scenarios we considered, MAIC was the least biased and most accurate method under no failures of assumptions, but robust sandwich standard errors underestimated variability when effective sample sizes were small, leading to undercoverage. Nevertheless, we recommend the use of MAIC for survival outcomes, provided that its assumptions are reasonable.

\item STC produced systematic bias because it targets a conditional treatment effect as opposed to a marginal treatment effect and the measure of effect is non-collapsible. The conditional measure of effect was incompatible in the indirect treatment comparison. We discourage the use of this version of STC, particularly when the measure of effect is non-collapsible. An important future objective is the development of an alternative formulation to STC that estimates a marginal treatment effect.

\item Future simulation studies should assess population adjustment methods with different outcome types and under model misspecification. 
\end{itemize}

\pagebreak

\begin{center}
	\tableofcontents
	\clearpage
	\listoftables
	\clearpage
	\listoffigures
\end{center}
\pagebreak


\renewcommand{\thefootnote}{\alph{footnote}}

\section{Introduction}\label{sec1}
\normalsize 

Evaluating the comparative effectiveness of alternative health care interventions lies at the heart of health technology assessments (HTAs), such as those commissioned by the National Institute of Health and Care Excellence (NICE), the body responsible for providing guidance on whether health care technologies should be publicly funded in England and Wales.\cite{sutton2008use} The randomized controlled trial (RCT) is the most reliable design for estimating the relative efficacy of new treatments.\cite{glenny2005indirect} However, new treatments are typically compared against placebo or standard of care before the licensing stage, but not necessarily against other active interventions --- a comparison that is required for HTAs. In the absence of data from head-to-head RCTs, indirect treatment comparisons (ITCs) are at the top of the hierarchy of evidence when assessing the relative efficacy of interventions and can inform treatment and reimbursement decisions.\cite{dias2013evidence}

Standard ITC techniques, such as network meta-analysis, are useful when there is a common comparator arm between RCTs, or more generally a connected network of studies.\cite{dias2013evidence, bucher1997results} These methods can be used with individual patient data (IPD) or aggregate-level data (ALD), with IPD considered the gold standard.\cite{stewart2002ipd} However, standard ITCs assume that there are no cross-trial differences in the distribution of effect-modifying variables (more specifically, that relative treatment effects are constant) and produce biased estimates when these exist.\cite{phillippo2018methods} Popular balancing methods such as propensity score matching\cite{austin2011introduction} can account for these differences but require access to IPD for all the studies being compared.\cite{faria2015nice} 

In many HTA processes, there are: (1) no head-to-head trials comparing the interventions of interest; (2) IPD available for at least one intervention (e.g.~from the submitting company’s own trial), but only published ALD for the relevant comparator(s); and (3) cross-trial differences in effect modifiers, implying that relative treatment effects are not constant across trial populations. Several methods, labeled \textit{population-adjusted indirect comparisons}, have been introduced to estimate relative treatment effects in this scenario. These include matching-adjusted indirect comparison (MAIC),\cite{signorovitch2010comparative, signorovitch2012comparative, signorovitch2012matching} based on inverse propensity score weighting,\cite{rosenbaum1987model} and simulated treatment comparison (STC),\cite{caro2010no} based on regression adjustment,\cite{zhang2009covariate} and require access to IPD from at least one of the trials.

The NICE Decision Support Unit has published formal submission guidelines for population adjustment with limited access to IPD.\cite{phillippo2018methods,phillippo2016nice} Various reviews\cite{phillippo2018methods,phillippo2016nice,ishak2015simulation,stevens2018review} define the relevant terminology and assess the theoretical validity of these methodologies but do not express a preference. Questions remain about the correct application of the methods and their validity in HTA.\cite{phillippo2018methods,phillippo2016nice,thom2016matching} Thus, Phillippo et al.\cite{phillippo2018methods} state that current guidance can only be provisional, as more thorough understanding of the properties of population-adjusted indirect comparisons is required. 

Consequently, several simulation studies have been published since the release of the NICE guidance.\cite{kuhnast2017evaluation, petto2019alternative, cheng2019statistical, hatswell2020effects, belger2015inclusion, leahy2019assessing} These have primarily assessed the performance of MAIC relative to standard~ITCs in a limited number of simulation scenarios. In general, the studies set relatively low effect modifier imbalances and do not vary these, even though MAIC is prone to large reductions in effective sample size and imprecise estimates of the treatment effect when high imbalances lead to poor overlap.\cite{phillippo2019population} Most importantly, existing simulation studies typically consider binary covariates at non-extreme values, not close to zero or one. In these scenarios, MAIC is likely to perform well as covariate overlap is strong. Propensity score weighting methods such as MAIC are known to be highly sensitive to scenarios with poor overlap,\cite{stuart2010matching, lee2011weight, hirano2001estimation} because of their inability to extrapolate beyond the observed covariate space. Hence, evaluating the performance of MAIC in the face of practical scenarios with poor covariate overlap is important. 

In this paper, we carry out an up-to-date review of MAIC and STC, and a comprehensive simulation study to benchmark the performance of the methods against the standard ITC. The simulation study provides proof-of-principle for the methods and is based on scenarios with survival outcomes and continuous covariates, with the log hazard ratio as the measure of effect. The methods are evaluated in a wide range of settings; varying the trial sample size, effect-modifying strength of covariates, prognostic effect of covariates, imbalance/overlap of covariates and the level of correlation in the covariates. 162 simulation scenarios are considered, providing the most extensive evaluation of population adjustment methods to date. An objective of the simulation study is to inform the circumstances under which population adjustment should be applied and which specific method is preferable in a given situation.

In Section \ref{sec2}, we establish the context and data requirements for population-adjusted indirect comparisons. In Section \ref{sec3}, we present an updated review of MAIC and STC. Section \ref{sec4} describes a simulation study, which evaluates the properties of these approaches under a variety of conditions. Section \ref{sec5} presents the results of the simulation study. An extended discussion of our findings and their implications is provided in Section \ref{sec6}. Finally, we make some concluding remarks in Section \ref{sec7}.

\section{Context}\label{sec2}

HTA often takes place late in the drug development process, after a new medical technology has obtained regulatory approval, typically based on a two-arm RCT that compares the new intervention to placebo or standard of care.  At the licensing stage, the question of interest is whether or not the drug is effective. In HTA, the relevant policy question is: ``given that there are finite resources available to finance health care, which is the best treatment of all available options in the market?''. In order to answer this question, one must evaluate the relative effectiveness of interventions that may not have been trialed against each~other. 

Indirect treatment comparison methods are used when we wish to compare the relative effect of interventions $A$ and $B$ for a specific outcome, but no head-to-head trials are currently available. Typically, it is assumed that the comparison is undertaken using additive effects for a given linear predictor, e.g.~log hazard ratio for time-to-event outcomes or log-odds ratio for binary outcomes. Indirect comparisons are typically performed on this scale.\cite{dias2013evidence, bucher1997results} In addition, we assume that the comparison is ``anchored'', i.e., a connected treatment network is available through a common comparator $C$, e.g.~placebo or standard of care. We note that comparisons can be unanchored, e.g.~using single-arm trials or disconnected treatment networks, but this requires much stronger assumptions.\cite{phillippo2018methods} The NICE Decision Support Unit discourages the use of unanchored comparisons when there is connected evidence and labels these as problematic.\cite{phillippo2018methods, phillippo2016nice} This is because they do not respect within-study randomization and are not protected from imbalances in any covariates that are prognostic of outcome (in essence implying that absolute outcomes can be predicted from the covariates, a heroic assumption). Hence, we do not present the methodology behind these. 

A manufacturer submitting evidence for reimbursement to HTA bodies has access to patient-level data from its own trial that compares its product $A$ against standard intervention $C$. However, as disclosure of proprietary, confidential patient-level data from industry-sponsored clinical trials is rare, IPD for the competitor's trial, comparing its treatment $B$ against $C$, are, almost invariably, unavailable (for both the manufacturer submitting evidence for reimbursement and the national HTA agency evaluating the evidence). We consider, without loss of generality, that IPD are available for a trial comparing intervention $A$ to intervention $C$ (denoted $AC$) and published ALD are available for a trial comparing $B$ to $C$ ($BC$). 

Standard methods for indirect comparisons such as the Bucher method,\cite{bucher1997results} a special case of network meta-analysis, allow for the use of ALD and estimate the $A$ vs.~$B$ treatment effect as:
\begin{equation}
\hat{\Delta}_{AB} = \hat{\Delta}_{AC} - \hat{\Delta}_{BC},
\label{equationn1}
\end{equation}
where $\hat{\Delta}_{AC}$ is the estimated relative treatment effect of $A$ vs.~$C$ (in the $AC$ population), and $\hat{\Delta}_{BC}$ is the estimated relative effect of $B$ vs.~$C$ (in the $BC$ population). The estimate $\hat{\Delta}_{AC}$ and an estimate of its variance can be calculated from the available IPD. The estimate $\hat{\Delta}_{BC}$ and an estimate of its variance may be directly published or derived from aggregate outcomes made available in the literature. As the indirect comparison is based on relative treatment effects observed in separate RCTs, the within-trial randomization of the originally assigned patient groups is preserved. The within-trial relative effects are statistically independent of each other; hence, their variances are simply summed to estimate the variance of the $A$ vs.~$B$ treatment effect.  

Standard indirect comparisons assume that there are no cross-trial differences in the distribution of effect-modifying variables. That is, the relative treatment effect of $A$ vs.~$C$ in the $AC$ population (indicated as $\Delta_{AC}$) is assumed equivalent to the treatment effect that would occur in the $BC$ population\footnote{In fact, standard ITC methods do not typically specify their target population explicitly (whether this is $AC$, $BC$ or otherwise), regardless of whether the analysis is based on ALD or on IPD from each study.\cite{manski2019meta}} (denoted $\Delta_{AC}^*$), --- throughout the paper the asterisk superscript represents a quantity that has been mapped to a different population; for example, in our case, the $A$ vs.~$C$ treatment effect in the $AC$ population is mapped to the population of the $BC$ trial.

Often, treatment effects are influenced by variables that interact with treatment on a specific scale (e.g.~the linear predictor), altering the effect of treatment on outcomes. If these \textit{effect modifiers} are distributed differently across $AC$ and $BC$, relative treatment effects differ in the trial populations and the assumptions of the Bucher method are broken. In this case, a standard ITC between $A$ and $B$ is liable to bias and may produce overly precise efficacy estimates.\cite{song2003validity} From the economic modeling point of view, these features are undesirable, as they impact negatively on the ``probabilistic sensitivity analysis'',\cite{baio2015probabilistic} the (often mandatory) process used to characterize the impact of the uncertainty in the model inputs on decision-making. 

As a result, population adjustment methodologies such as MAIC and STC have been introduced. These target the $A$ vs.~$C$ treatment effect that would be observed in the $BC$ population, thereby performing an adjusted indirect comparison in such population. The adjusted $A$ vs.~$B$ treatment effect is estimated as:
\begin{equation}
\hat{\Delta}_{AB}^* = \hat{\Delta}_{AC}^* - \hat{\Delta}_{BC},
\label{equationn2}
\end{equation}
where $\hat{\Delta}_{AC}^*$ is the estimated relative treatment effect of $A$ vs $C$ (in the $BC$ population, implicitly assumed to be the relevant target population). Variances are combined in the same way as the Bucher method. 

Those studying the generalizability of treatment effects often make a distinction between sample and population treatment effects.\cite{stuart2011use, cole2010generalizing, kern2016assessing, hartman2015sample} Typically, another implicit assumption made by population-adjusted indirect comparisons is that the treatment effects estimated in the $BC$ sample, as described by its published covariate moments in the case of $\hat{\Delta}_{AC}^{*}$, coincide with those that would be estimated in the target population of the trial. Namely, either the study sample on which inferences are made is the study target population, or it is a simple random sample (i.e., representative) of such population, ignoring the sampling variability in the descriptive characteristics.

\begin{figure}[!htb]
\center{\includegraphics[width=\textwidth]{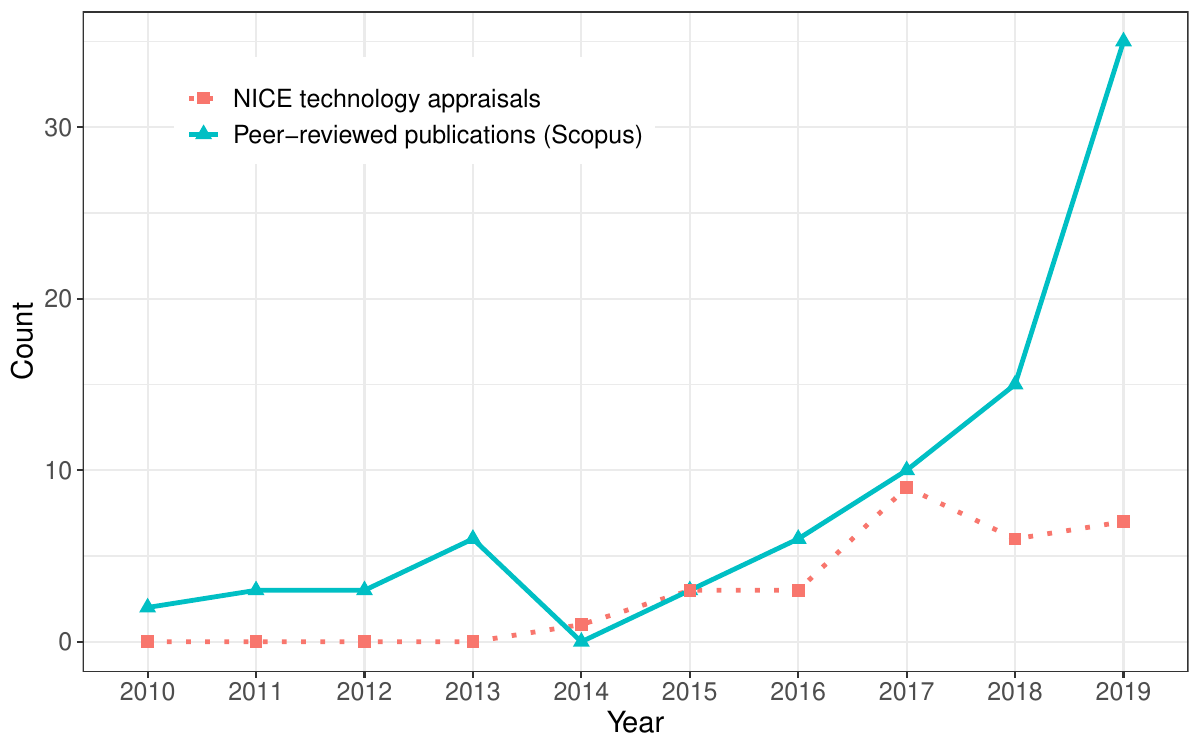}}
\caption{Number of peer-reviewed publications and technology appraisals from the National Institute for Health and Care Excellence (NICE) using population-adjusted indirect comparisons per year.\label{fig1}}
\end{figure}

\clearpage

The use of population adjustment in HTA, both in published literature as well as in submissions for reimbursement, and its acceptability by national HTA bodies, e.g.~in England and Wales, Scotland, Canada and Australia,\cite{thom2016matching} is increasing across diverse therapeutic areas.\cite{thom2016matching, phillippo2019population, veroniki2016scoping, ndirangu2016trends} As of April 11, 2020, a search among titles, abstracts and keywords for ``matching-adjusted indirect comparison'' and ``simulated treatment comparison'' in Scopus, reveals at least 89 peer-reviewed applications of MAIC and STC and conceptual papers about the methods. In addition, at least 30 technology appraisals (TAs) published by NICE use MAIC or STC --- of these, 23 have been published since 2017. Figure \ref{fig1} shows the rapid growth of peer-reviewed publications and NICE TAs featuring MAIC or STC since the introduction of these methods in 2010. MAIC and STC are predominantly applied in the evaluation of cancer drugs, as 26 of the 30 NICE TAs using population adjustment have been in oncology.

\section{Methodology}\label{sec3}

We shall assume that the following data are available for the $i-$th subject ($i=1, \dots, N$) in the $AC$ trial:
\begin{itemize}
\item A covariate vector of $K$ baseline characteristics $\boldsymbol{X}_i=(X_{i,1},\dots,X_{i,K})$, e.g.~age, gender, comorbidities;
\item A treatment indicator $T_i$. Without loss of generality, we assume here for simplicity that $T_i$ $\in \{ 0, 1\}$ for the common comparator and active treatment, respectively;
\item An observed outcome $Y_i$, e.g.~a time-to-event or binary indicator for some clinical measurement.
\end{itemize}
Given this information, one can compute an unadjusted estimate $\hat{\Delta}_{AC}$ of the $A$ vs.~$C$ treatment effect, and an estimate of its variance. In the Bucher method, such estimate would be plugged in to Equation \ref{equationn1}. On the other hand, MAIC and STC generate a population-adjusted estimate $\hat{\Delta}_{AC}^*$ of the $A$ vs.~$C$ treatment effect that would be plugged in to Equation \ref{equationn2}.

For the $BC$ trial, data available are:
\begin{itemize}
\item A vector $\bar{\boldsymbol{X}}_{BC} = (\bar{X}_{BC, 1}, \dots,  \bar{X}_{BC, K})$ of published summary values for the baseline characteristics. For ease of exposition, we shall assume that these are means and are available for all $K$ covariates (alternatively, one would take the intersection of the available covariates).
\item An estimate $\hat{\Delta}_{BC}$ of the $B$ vs.~$C$ treatment effect in the $BC$ population, and an estimate of its variance, either published directly or derived from aggregate outcomes in the literature. 
\end{itemize}
Each baseline characteristic $k=1,\dots,K$ can be classed as a prognostic variable (a covariate that affects outcome), an effect modifier (a covariate that interacts with treatment $A$ to affect outcome), both or none. For simplicity in the notation, it is assumed that all available baseline characteristics are prognostic of the outcome and that a subset of these, $\boldsymbol{X}_i^{\boldsymbol{(EM)}} \subset \boldsymbol{X}_i$, are selected as effect modifiers (of treatment $A$) on the linear predictor scale. Similarly, for the published summary values, $\bar{\boldsymbol{X}}^{(EM)}_{BC} \subset 
\bar{\boldsymbol{X}}_{BC}$. Note that we select the effect modifiers of treatment $A$ with respect to $C$ (as opposed to the effect modifiers of treatment $B$ with respect to $C$), because we have to adjust for these to perform the indirect comparison in the $BC$ population, implicitly assumed to be the target population.\footnote{If we had IPD for the $BC$ study and ALD for the $AC$ study, we would have to adjust for the covariates that modify the effect of treatment $B$ vs.~$C$, in order to perform the comparison in the $AC$ population.} 

\subsection{Matching-adjusted indirect comparison}\label{subsec31}

Matching-adjusted indirect comparison (MAIC) is a population adjustment method based on inverse propensity score weighting.\cite{rosenbaum1987model} IPD from the $AC$ trial are weighted so that the means and, potentially, higher moments of specified covariates match those in the $BC$ trial. The weights are estimated using a propensity score logistic regression model:
\begin{equation*}
\ln{(w_{i})} =  \alpha_0 + \boldsymbol{X}_i^{\boldsymbol{(EM)}} \boldsymbol{\alpha_1},
\end{equation*}
where $\alpha_0$ and $\boldsymbol{\alpha_1}$ are the regression parameters, and the weight $w_i$ assigned to each individual $i$ represents the ``trial selection'' odds, i.e., the odds of being enrolled in the $BC$ trial as opposed to being enrolled in the $AC$ trial. These are defined as a function of the baseline characteristics modifying the effect of treatment $A$,  $\boldsymbol{X}_i^{\boldsymbol{(EM)}}$ for subject $i$. Note that in standard applications of propensity score weighting, e.g.~in observational studies, the propensity score logistic regression is for the \textit{treatment group} assigned to the subject. In MAIC, the objective is to balance covariates across studies so the propensity score model is for the \textit{trial} in which the participant is enrolled. 

The regression parameters cannot be derived using conventional methods such as maximum-likelihood estimation because IPD are not available for $BC$. Signorovitch et al.\cite{signorovitch2010comparative} propose using a method of moments to estimate the model parameters by setting the weights so that the mean effect modifiers are exactly balanced across the two trial populations. After centering the $AC$ effect modifiers on the published $BC$ means, such that $\bar{\boldsymbol{X}}_{BC}^{(EM)} = \boldsymbol{0}$, the weights are estimated by minimizing the objective function:
\begin{equation*}
Q(\boldsymbol{\alpha}_1) = \sum_{i=1}^{N} \exp \left ( \boldsymbol{X}_i^{\boldsymbol{(EM)}}\boldsymbol{\alpha_1} \right ),
\end{equation*}
where $N$ represents the number of subjects in the $AC$ trial. $Q(\boldsymbol{\alpha}_1)$ is a convex function that can be minimized using standard algorithms, e.g.~BFGS,\cite{nocedal2006numerical} to yield a unique finite solution $\hat{\boldsymbol{\alpha}}_{\boldsymbol{1}}=\textnormal{argmin}(Q(\boldsymbol{\alpha}_1))$. Then, the estimated weight for subject $i$ is:
\begin{equation*}
\hat{w}_i = \exp(\boldsymbol{X}_i^{\boldsymbol{(EM)}}\boldsymbol{\hat{\alpha}}_{\boldsymbol{1}}).
\end{equation*}
Consequently, the mean outcomes under treatment $t \in \{A,C\}$ in the $BC$ population are predicted as the weighted average:
\begin{equation*}
\hat{Y}_t^* = \frac{\sum_{i=1}^{N_t} Y_{i,t} \hat{w}_{i}}{\sum_{i=1}^{N_t} \hat{w}_{i}},
\label{eq2}
\end{equation*}
where $N_t$ represents the number of subjects in arm $t$ of the $AC$ trial, and $Y_{i,t}$ denotes the outcome for patient $i$ receiving treatment $t$ in the patient-level data. Note that we have summary data from the $BC$ trial to estimate absolute outcomes under $C$. However, in the anchored scenario, we do not focus on the absolute outcomes as the objective is to generate a relative effect for $A$ vs.~$C$ in the $BC$ population.

Such relative effect is typically estimated by fitting a weighted model, i.e., a model where the contribution of each subject to the likelihood is weighted. For instance, if the outcome of interest is a time-to-event outcome, an ``inverse odds''-weighted Cox model can be fitted by maximizing its weighted partial likelihood. In this case, a subject $i$ from the $AC$ trial, who has experienced an event at time $\tau$, contributes the following term to the partial likelihood function:
\begin{equation}
\Bigg (
\frac{\exp(\beta_TT_i)}
{\sum_{j \in R(\tau)} \hat{w}_j \exp(\beta_TT_j)}
\Bigg )^{\hat{w}_i},
\label{eqn21}
\end{equation}
where $R(\tau)$ is the set of subjects without the event and uncensored prior to $\tau$, i.e., the risk set. Here, the fitted coefficient $\hat{\beta}_T$ of the weighted regression (i.e., the value of the parameter maximizing the partial likelihood in Equation \ref{eqn21}) is the estimated relative effect for $A$ vs.~$C$, such that $\hat{\Delta}_{AC}^*=\hat{\beta}_T$. 

In the original MAIC approach, covariates are balanced for active treatment and control arms combined and standard errors are computed using a robust sandwich estimator, which allows for heteroskedasticity.\cite{signorovitch2010comparative, white1980heteroskedasticity} Typically, implementations of this estimator do not explicitly account for the fitting of the logistic regression model for the weights, assuming these to be fixed. 

Terms of higher order than means can also be balanced, e.g.~by including squared covariates in the method of moments to match variances. However, this decreases the degrees of freedom and may increase finite-sample bias.\cite{windmeijer2005finite} Matching both means and variances (as opposed to means only) appears to result in more biased and less accurate treatment effect estimates when covariate variances differ across trials.\cite{petto2019alternative, hatswell2020effects}

A proposed modification to MAIC uses entropy balancing\cite{hainmueller2012entropy} instead of the method of moments to estimate the weights.\cite{petto2019alternative, belger2015inclusion} Entropy balancing has the additional constraint that the weights are as close as possible to unit weights. Potentially, it should penalize extreme weighting schemes and provide greater precision. However, Phillippo et al. recently demonstrated that weight estimation via entropy balancing and the method of moments are mathematically identical.\cite{phillippo2020equivalence} Other proposed modifications to MAIC include balancing the covariates separately for active treatment and common comparator arms,\cite{petto2019alternative, belger2015inclusion} and using the bootstrap\cite{efron1992bootstrap, efron1994introduction} to compute standard errors,\cite{sikirica2013comparative} which does not rely upon strong assumptions about the estimation of the MAIC weights. Balancing the covariates separately seems to provide greater precision in simulation studies.\cite{petto2019alternative} However, we do not recommend this approach because it may break randomization, distorting the balance between treatment arms $A$ and $C$ on covariates that are not accounted for in the weighting. If these covariates are prognostic of outcome, this would compromise the internal validity of the within-study treatment effect estimate for $A$ vs.~$C$. 

As MAIC is a reweighting procedure, it will reduce the effective sample size (ESS) of the $AC$ trial. The approximate ESS of the weighted IPD is estimated as $\left(\sum_i\hat{w}_i\right)^2/\sum_i\hat{w}_i^2$; the reduction in ESS can be viewed as a rough indicator of the lack of overlap between the $AC$ and $BC$ covariate distributions. For relative effects to be conditionally constant and eventually produce an unbiased indirect comparison, one needs to include all effect modifiers in the weighting procedure, whether in imbalance or not (see \hyperref[SA]{Supplementary Appendix A} for a non-technical overview of the full set of assumptions made by MAIC, and more generally, by population-adjusted indirect comparisons).\cite{phillippo2016nice} The exclusion of balanced covariates does not ensure their balance after the weighting procedure. Including too many covariates or poor overlap in the covariate distributions can induce extreme weights and large reductions in ESS. This is a pervasive problem in NICE TAs, where most of the reported ESSs are small with a large percentage reduction from the original sample size.\cite{phillippo2019population} 

Propensity score mechanisms are very sensitive to poor overlap.\cite{stuart2010matching, lee2011weight, hirano2001estimation} In particular, weighting methods are unable to extrapolate --- in the case of MAIC, extrapolation beyond the covariate space observed in the $AC$ IPD is not possible. Almost invariably, the level of overlap between the covariate distributions will decrease as a greater number of covariates are included. Therefore, no purely prognostic variables should be balanced to avoid loss of effective sample size and consequent inflation of the standard error due to over-balancing.\cite{phillippo2018methods} Cross-trial imbalances in purely prognostic variables should not produce bias as relative treatment effects are unaffected in expectation due to within-trial randomization.\cite{phillippo2016nice} 

\subsection{Simulated treatment comparison}\label{subsec32}

While MAIC is a reweighting method, simulated treatment comparison (STC)\cite{caro2010no} is based on regression adjustment.\cite{zhang2009covariate} Regression adjustment methods are promising because they may increase precision and statistical power with respect to propensity score-based methodologies.\cite{steyerberg2019clinical, harrell2016biostatistics,senn2007stratification} Contrary to most propensity score methods, regression adjustment mechanisms are able to extrapolate beyond the covariate space where overlap is insufficient, using the linearity assumption or other appropriate assumptions about the input space. However, the validity of the extrapolation depends on the accuracy in capturing the true covariate-outcome relationships.   

In the typical version of STC, IPD from the $AC$ trial are used to fit a regression of the outcome on the baseline characteristics and treatment. Following the NICE Decision Support Unit Technical Support Document 18,\cite{phillippo2018methods,phillippo2016nice} the following linear predictor is fitted to the IPD:
\begin{equation}
g(\eta^*_i) = \beta_0 +  \left(\boldsymbol{X}_i  - \bar{\boldsymbol{X}}_{BC}  \right)\boldsymbol{\beta_1} + \left[\beta_T +  \left(\boldsymbol{X}_i^{\boldsymbol{(EM)}} - \bar{\boldsymbol{X}}_{BC}^{\boldsymbol{(EM)}} \right) \boldsymbol{\beta_2}\right]\mathds{1}(T_i=1), 
\label{eq3}
\end{equation}
where $\eta^*_i$ is the expected outcome on the natural outcome scale, e.g.~the probability scale for binary outcomes, of subject $i$, $g(\cdot)$ is an appropriate link function (e.g.~logit for binary outcomes), $\beta_0$ is the intercept, $\boldsymbol{\beta_1}$ is a vector of $K$ regression coefficients for the prognostic variables, $\boldsymbol{\beta_2}$ is a vector of interaction coefficients for the effect modifiers (modifying the effect of treatment $A$ vs.~$C$) and $\beta_T$ is the $A$ vs.~$C$ treatment coefficient. The covariates are centered at the published mean values from the $BC$ population, $\bar{\boldsymbol{X}}_{BC}$ and $\bar{\boldsymbol{X}}_{BC}^{\boldsymbol{(EM)}}$, respectively. Hence, the estimated $\hat{\beta}_T$ is directly interpreted as the $A$ vs.~$C$ treatment effect in the $BC$ population, such that $\hat{\Delta}_{AC}^*=\hat{\beta}_T$. The variance of said treatment effect is derived directly from the fitted model (see\cite{phillippo2018methods,phillippo2016nice} for a breakdown of uncertainty propagation in the estimates resulting from MAIC and STC). In a Cox proportional hazards regression framework, a log link function could be employed in Equation \ref{eq3} between the hazard function and the linear predictor component of the model. 

For relative effects to be conditionally constant across studies, one needs to include in the model the effect modifiers that are imbalanced. In addition, the relationship between the effect modifiers and outcome must be correctly specified; in the case of this article, the effect modifiers must have an additive interaction with treatment on the linear predictor scale. It is optional to include (and to center) imbalanced variables that are purely prognostic. These will not remove bias further but a strong fit of the outcome model may increase precision. NICE guidance\cite{phillippo2016nice} suggests adding purely prognostic variables if they increase the precision of the model and account for more of its underlying variance, as reported by model selection criteria (e.g.~residual deviance or information criteria). However, such tools should not guide decisions on effect modifier status, which must be defined prior to fitting the outcome model. As effect-modifying covariates are likely to be good predictors of outcome, the inclusion of appropriate effect modifiers should provide an acceptable fit.

Alternative ``simulation-based'' formulations to STC have been proposed.\cite{ishak2015simulation, ishak2015simulated} These are outlined as follows. The joint distribution of $BC$ covariates is approximated under certain parametric assumptions to characterize the $BC$ population, e.g.~simulating continuous covariates at the individual level from a multivariate normal with the $BC$ means and the correlation structure observed in the $AC$ IPD. A regression of the outcome on the predictors is fitted to the $AC$ patient-level data (this time, the covariates are not centered at the mean $BC$ values). Then, the coefficients of this regression are applied to the simulated subject profiles and the linear predictions for patients under $A$ and under $C$ in the $BC$ population are averaged out. The treatment effect for $A$ vs.~$C$ is given by subtracting the average linear prediction under $C$ from the average linear prediction under $A$. Neither the original conceptual publications nor NICE guidance provide detailed information about variance estimation, which is likely to be complicated and probably requires bootstrapping or similar approaches. 

It is worth noting that, in the linear predictor scale, the arithmetic mean of the average linear predictor (the average linear predictor for patients sampled under the centered covariates) and its geometric mean (the linear predictor evaluated at the expectation of the centered covariates) coincide. Therefore, provided that the number of simulated subjects is sufficiently large (i.e., in expectation or ignoring sampling variability), the ``covariate simulation'' approach generates estimates that are equivalent to those of the ``plug-in'' methodology adopted in this article. 


\subsection{Clarification of estimands}\label{subsec33}

In an indirect treatment comparison, the objective is to emulate the analysis of a head-to-head RCT between $A$ and $B$. However, RCTs have two potential target estimands: \textit{marginal} and/or \textit{conditional} treatment effects. 
In MAIC, as is typically the case for propensity score methods, $\hat{\Delta}_{AC}^*$ targets a \textit{marginal} treatment effect.\cite{austin2011introduction, joffe2004model, rosenbaum1998encyclopedia} In biostatistics\cite{austin2013performance, austin2010performance, austin2010substantial,neuhaus1991comparison} and epidemiology,\cite{greenland1987interpretation, hubbard2010gee, hernan2004definition} this marginal effect is also known as a \textit{population-average} or \textit{population-level} treatment effect, as it measures the average treatment effect for $A$ vs.~$C$ at the population level (conditional on the entire population distribution of covariates, such that the individual-level covariates have been marginalized over). This denotes the average outcome between two identical populations, except that in one population all subjects are under $A$, while in the other population all subjects are under $C$,\cite{austin2014use} and where the difference is taken on a suitable transformed scale, e.g.~the linear predictor scale. MAIC targets a marginal treatment effect because it performs a weighted regression of outcome on treatment assignment alone. Therefore, assuming a reasonably large sample size and proper randomization in the $AC$ trial, the fitted coefficient $\hat{\beta}_T$ in Equation \ref{eqn21} estimates a relative effect between subjects that have the same distribution of baseline characteristics (corresponding to the $BC$ population). 

In HTA and health policy, interest typically lies in the impact of a health technology on the target population for the decision problem, which MAIC and STC implicitly assume to be the $BC$ population. Where making decisions at the population level, the effect of interest is a marginal treatment effect: the average effect, at the population level, of moving the target population from treatment $B$ to treatment $A$.\cite{austin2011introduction, imbens2004nonparametric} The majority of trials report an estimate $\hat{\Delta}_{BC}$ that targets a marginal treatment effect. It is likely derived from a RCT publication where a simple regression of outcome on a single independent variable, treatment assignment, has been fitted. 

In the version of STC outlined by the NICE Decision Support Unit, $\hat{\Delta}_{AC}^*$ targets a \textit{conditional} treatment effect. The conditional treatment effect denotes the average effect, at the individual level, of changing a subject's treatment from $C$ to $A$.\cite{austin2011introduction,austin2014use} STC targets a conditional treatment effect because the estimate is obtained from the regression coefficient of a multivariable regression ($\hat{\beta}_T$ in Equation \ref{eq3}), where the baseline covariates included as predictors are also adjusted for. Hence, the relative effect is an average at the subject level, fully conditioned on the covariates of the average subject.\footnote{While the treatment coefficient $\hat{\beta}_T$ is an \textit{average} treatment effect, it is not a population-level measure, contrary to the \textit{marginal} or \textit{population-average} treatment effect, which is the effect of moving the entire population from one treatment to the other. Firstly, while there is only \textit{one} marginal effect for a specific population (as described by its covariate distribution), there may be many average conditional effects for a given population, one for every possible combination of covariates and model specification considered for adjustment. Secondly, conditioning on covariates changes the nature of the effect, so that it is no longer a population-level effect but an effect with a subject-level interpretation. Note that, with non-collapsible effect measures, marginal effects may not be equal to a weighted average of the conditional effects under any weighting scheme, even under no confounding bias.\cite{huitfeldt2019collapsibility}} Conditional measures of effect are clinically relevant as patient-centered evidence in a clinician–patient context, where decision-making relates to the treatment benefit for an individual subject with specific covariate values. Conditional treatment effects are typically not of interest when making decisions at the population level in HTA and health policy, as they are unit-level measures of effect. 

A measure of effect is said to be \textit{collapsible} if marginal and conditional effects coincide in the absence of confounding bias. \cite{greenland1987interpretation, greenland1999confounding} The property of collapsibility is closely related to that of linearity,\cite{janes2010quantifying, martinussen2013collapsibility} e.g.~mean differences in a linear regression are collapsible.\cite{austin2011introduction, austin2014use, greenland1987interpretation, greenland1999confounding} However, most applications of population-adjusted indirect comparisons are in oncology and are typically concerned with time-to-event outcomes, or rate outcomes modeled using logistic regression.\cite{phillippo2019population} These yield non-collapsible measures of treatment effect such as (log) hazard ratios\cite{austin2011introduction, austin2014use, greenland1987interpretation, gail1984biased} or (log) odds ratios.\cite{austin2011introduction, austin2014use, greenland1987interpretation,greenland1999confounding, gail1984biased, miettinen1981confounding, neuhaus1991comparison} 

With non-collapsible measures of effect, marginal and conditional estimands do not coincide due to non-linearity,\cite{janes2010quantifying} even if there is covariate balance and no confounding.\cite{greenland1987interpretation, greenland1999confounding} With both collapsible and non-collapsible measures of effect, estimators targeting distinct estimands will have different standard errors. Therefore, marginal and conditional estimates quantify parametric uncertainty differently, and conflating these will lead to the incorrect propagation of uncertainty to the wider health economic decision model, which will be problematic for probabilistic sensitivity analyses.  

Therefore, the relative effect estimate $\hat{\Delta}_{AC}^{(*)}$ in STC is unable to target a marginal treatment effect and the comparison of interest, a comparison of compatible marginal effects, cannot be performed. A comparison of conditional effects is not of interest, and also, cannot be carried out. A compatible conditional effect for $B$ vs.~$C$ is unavailable because its estimation requires fitting the non-centered version of Equation \ref{eq3}, adjusting for the same baseline characteristics, to the $BC$ patient-level data. Such data are unavailable and it is unlikely that the estimated treatment coefficient from this model is available in the clinical trial publication. 

Hence, $\hat{\Delta}_{AC}^*$ is incompatible with $\hat{\Delta}_{BC}$ in the indirect comparison (Equation \ref{equationn2}) for STC, even if all effect modifiers are accounted for and the outcome model is correctly specified. If we intend to target a marginal estimand for the $A$ vs.~$C$ treatment effect (in the $BC$ population) and naively assume that STC does so, $\hat{\Delta}_{AB}^*$ may produce a biased estimate of the marginal treatment effect for $A$ vs.~$B$, even if all the assumptions in \hyperref[SA]{Supplementary Appendix A} are met. On the other hand, $\hat{\Delta}_{AC}^*$ targets a marginal treatment effect in MAIC. There are no compatibility issues in the indirect comparison as $\hat{\Delta}_{AC}^*$ and $\hat{\Delta}_{BC}$ target compatible estimands of the same form. In the Bucher method, if the estimate $\hat{\Delta}_{AC}$ is derived from a simple comparison of group means or from an univariable regression of outcome on treatment in the $AC$ IPD, this targets a marginal effect and there are no compatibility issues in the indirect treatment comparison either.

\section{Simulation Study}\label{sec4}

\renewcommand{\thefootnote}{\alph{footnote}}

\subsection{Aims}\label{subsec41}

The objectives of the simulation study are to compare MAIC, STC and the Bucher method across a wide range of scenarios that may be encountered in practice. For each estimator, we assess the following properties:\cite{morris2019using} (1) unbiasedness; (2) variance unbiasedness; (3) randomization validity;\footnote{In a sufficiently large number of repetitions, $(100\times(1-\alpha))$\% confidence intervals based on normal distributions should contain the true value $(100\times(1-\alpha))$\% of the time, for a nominal significance level $\alpha$.} and (4) precision. The selected performance measures evaluate these criteria specifically (see \ref{subsec45}). The simulation study is reported following the ADEMP (Aims, Data-generating mechanisms, Estimands, Methods, Performance measures) structure.\cite{morris2019using} All simulations\footnote{The files required to run the simulations are available at \url{http://github.com/remiroazocar/population_adjustment_simstudy}. \hyperref[SB]{Supplementary Appendix B} lists the specific settings of each simulation scenario.}  and analyses were performed using \texttt{R} software version 3.6.3.\cite{team2013r} Example \texttt{R} code implementing MAIC, STC and the Bucher method on a simulated example is provided in \hyperref[SD]{Supplementary Appendix D}.

\subsection{Data-generating mechanisms}\label{subsec42}

As most applications of MAIC and STC are in oncology, the most prevalent outcome types are survival or time-to-event outcomes (e.g.~overall or progression-free survival).\cite{phillippo2019population} Hence we consider these using the log hazard ratio as the measure of effect. 

\begin{figure}[!htb]
\center{\includegraphics[width=0.82\textwidth]{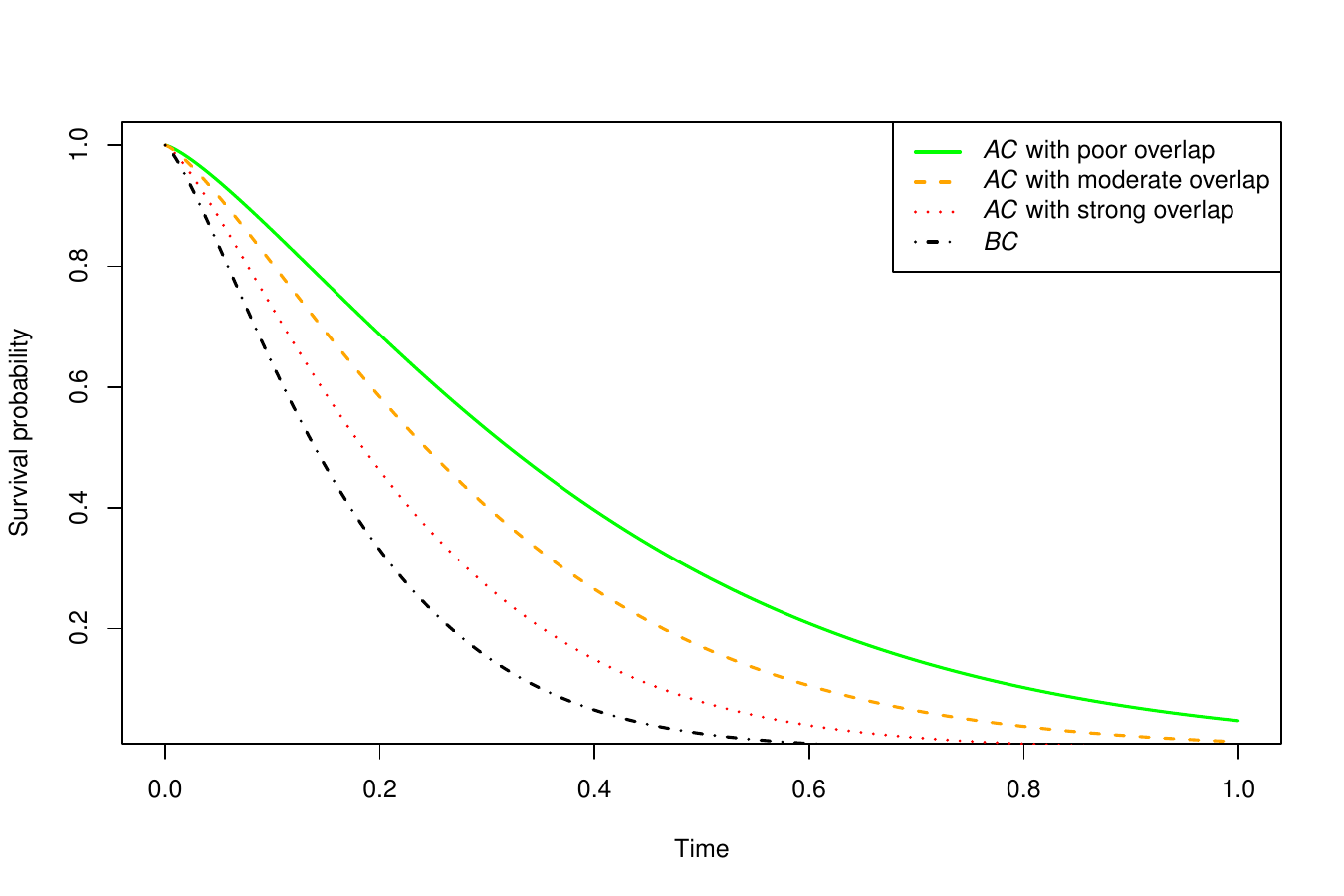}}
\caption{Weibull-distributed curves used to simulate survival times for subjects under the active treatment for different trial populations. The covariates are associated with shorter survival and, in the case of the effect modifiers, interact with treatment to render it less effective. As the mean values of the $AC$ covariates decrease, overlap decreases.\label{figweib1}}
\end{figure}

\begin{figure}[!htb]
\center{\includegraphics[width=0.82\textwidth]{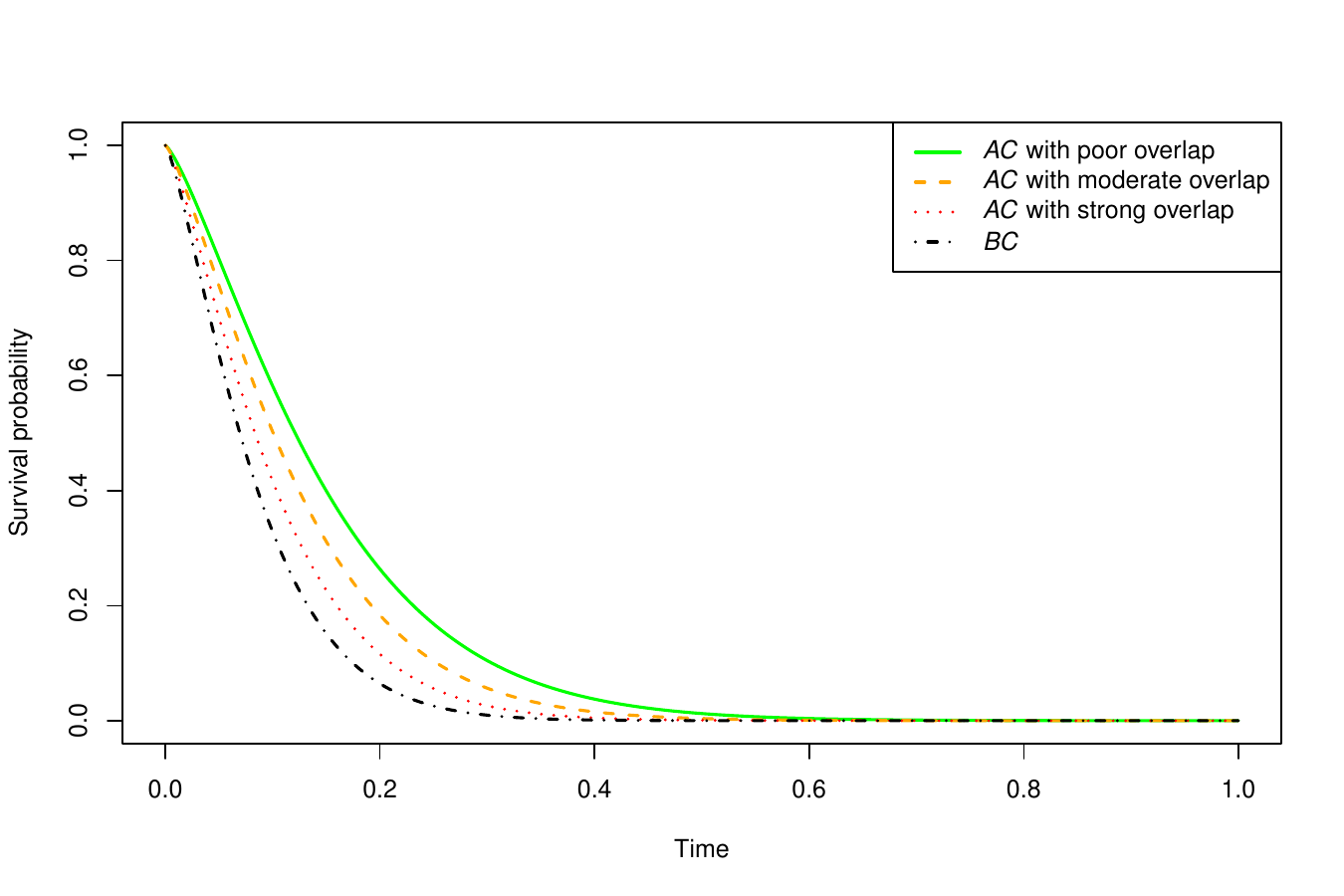}}
\caption{Weibull-distributed curves used to simulate survival times for subjects under the common comparator for different trial populations.\label{figweib2}}
\end{figure}

For trials $AC$ and $BC$, we follow Bender et al.\cite{bender2005generating} to simulate Weibull-distributed survival times under a proportional hazards parametrization.\footnote{At baseline, this formulation has a hazard function $h_0(\tau)=\lambda \nu \tau^{\nu-1}$, a cumulative hazard function $H_0(\tau)=\lambda \tau^\nu$, a density function $f_0(\tau) = \lambda \nu \tau^{\nu-1} \exp(-\lambda \tau^\nu)$ and a survival function $S_0(\tau) = \exp(-\lambda \tau^\nu)$ at time $0 \leq \tau < \infty$, where $\lambda>0$ is a positive inverse scale (rate) parameter, and $\nu >0$ is a positive shape parameter. This follows the proportional hazards parametrization of the Weibull distribution in NICE guidelines, where $\lambda$ is referred to as a scale parameter.\cite{latimer2013survival}} Survival time $\tau_i$ (for subject $i$) is generated according to the formula:
\begin{equation}
\tau_i = \Bigg ( \frac{-\ln U_i}{\lambda \exp 
[\boldsymbol{X}_i \boldsymbol{\beta_1} + (\beta_T +  \boldsymbol{X}_i^{\boldsymbol{(EM)}} \boldsymbol{\beta_2})\mathds{1}(T_i = 1)]} \Bigg )^{1/\nu},
\label{eq4}
\end{equation}
where $U_i$ is a uniformly distributed random variable, $U_i \sim (0, 1)$. We set the inverse scale of the Weibull distribution to $\lambda=8.5$ and the shape to $\nu=1.3$ as these parameters produce a functional form reflecting frequently observed mortality trends in metastatic cancer patients (as illustrated in Figure \ref{figweib1} and Figure \ref{figweib2}, which display the survival curves implied by the parameters).\cite{hatswell2020effects} Four correlated or uncorrelated continuous covariates $\boldsymbol{X}_i$ are generated per subject using a multivariate Gaussian copula.\cite{nelsen2007introduction} Two of these are purely prognostic variables; the other two ($\boldsymbol{X}_i^{\boldsymbol{(EM)}}$) are effect modifiers, modifying the effect of both treatments $A$ and $B$ with respect to $C$ on the log hazard ratio scale, and prognostic variables. 

We introduce random right censoring to simulate loss to follow-up within each trial. Censoring times $\tau_{c,i}$ are generated from the exponential distribution $\tau_{c,i} \sim \textnormal{Exp}(\lambda_c)$, where the rate parameter $\lambda_c=0.96$ is selected to achieve a censoring rate of 35\% under the active treatment at baseline (with the values of the covariates set to zero), considered moderate censoring.\cite{abrahamowicz2004bias} We fix the value of $\lambda_c$ before generating the datasets, by simulating survival times for 1,000,000 subjects with Equation \ref{eq4} and using the \texttt{R} function \texttt{optim} (Brent's method\cite{brent1971algorithm}) to minimize the difference between the observed and targeted censoring proportion.

The number of subjects in the $BC$ trial is 600, under a 1:1 active treatment vs.~control allocation ratio. This sample size corresponds to that of a reasonably large Phase III RCT \cite{stanley2007design}. Different values are not explored as preliminary results showed that these drive performance less than the number of subjects in the $AC$ trial. While the number of subjects in $BC$ contributes to sampling variability, the reweighting or regressions are performed in the $AC$ patient-level data. For the $BC$ trial, the individual-level covariates and outcomes are aggregated to obtain summaries. The continuous covariates are summarized as means --- these would typically be available to the analyst in the published study as a table of baseline characteristics. The marginal $B$ vs.~$C$ treatment effect and its variance are estimated through a Cox proportional hazards regression of outcome on treatment. These estimates make up the only information on aggregate outcomes available to the analyst. 

The simulation study examines five factors in a fully factorial arrangement with $3 \times 3 \times 3 \times 2 \times 3 = 162$ scenarios to explore the interaction between factors. The simulation scenarios are defined by varying the values of the following parameters, which are inspired by applications of MAIC and STC in NICE technology appraisals:

\begin{itemize}
\item The number of patients in the $AC$ trial, $N \in \{150, 300, 600\}$ under a 1:1 active intervention vs.~control allocation ratio. The sample sizes correspond to typical values for a Phase III RCT\cite{stanley2007design} and for trials included in applications of MAIC and STC submitted to HTA authorities.\cite{phillippo2019population}
\item The strength of the association between the prognostic variables and the outcome, $\beta_{1, k} \in \{ -\ln(0.67), -\ln(0.5), \allowbreak -\ln(0.33) \}$ (moderate, strong and very strong prognostic variable effect), where $k$ indexes a given covariate. These regression coefficients correspond to fixing the conditional hazard ratios for the effect of each prognostic variable at approximately 1.5, 2 and approximately 3, respectively. 
\item The strength of interaction of the effect modifiers, $\beta_{2,k} \in \{ -\ln(0.67), - \ln(0.5), \allowbreak -\ln(0.33) \}$ (moderate, strong and very strong interaction effect), where $k$ indexes a given effect modifier. These parameters have a material impact on the marginal $A$ vs.~$B$ treatment effect. Hence, population adjustment is warranted in order to remove the induced bias.
\item The level of correlation between covariates, $\textnormal{cor}(X_{i,k}, X_{i,l})  \in \{0,0.35\}$ (no correlation and moderate correlation), for subject $i$ and covariates $k \neq l$.
\item The degree of covariate imbalance.\footnote{Due to the simulation study design, where the covariate distributions are symmetric, covariate \textit{balance} is a proxy for covariate \textit{overlap} in this parameter setting. Imbalance refers to the difference in covariate distributions across studies, as measured by the difference in (standardized) average covariate values. Overlap describes the degree of similarity in the covariate ranges across studies --- there is complete overlap if the ranges are the same. In real scenarios, lack of complete overlap does not necessarily imply imbalance (and vice versa). Imbalances in effect modifiers across studies bias the standard indirect comparison, motivating the use of population adjustment. Lack of complete overlap hinders the use of population adjustment, as the covariate data may be too limited to make any conclusions in the regions of non-overlap.} For both trials, each covariate $k$ follows a normal marginal distribution. For the $BC$ trial, we fix $X_{i,k} \sim \textnormal{Normal}(0.6, 0.2^2)$, for subject $i$. For the $AC$ trial, the normal distributions have mean $\mu_k$, such that $X_{i,k} \sim \textnormal{Normal}(\mu_k, 0.2^2)$, varying $\mu_k \in \{ 0.45, 0.3, 0.15\}$. This yields strong, moderate and poor covariate overlap, respectively, corresponding to average percentage reductions in ESS across scenarios of 19\%, 53\% and 79\%. These percentage reductions in ESS are representative of the range encountered in NICE TAs (see below).
\end{itemize}

Each active intervention has a very strong conditional treatment effect $\beta_T = \ln(0.25)$ versus the common comparator. The covariates may represent comorbidities, which are associated with shorter survival and, in the case of the effect modifiers, which interact with treatment to render it less effective. Figure \ref{figweib1} shows the Weibull-distributed survival curves for patients under the active treatment ($A$ and $B$) with varying levels of the covariates. Figure \ref{figweib2} shows the Weibull-distributed survival curves for subjects under the common comparator ($C$). In Figures \ref{figweib1} and \ref{figweib2}, the strength of each prognostic term and each effect-modifying interaction is moderate.

The varying degrees of covariate overlap are inspired by applications of MAIC in technology appraisals submitted to the NICE. Only 13 of the 27 appraisals carrying out a MAIC have effective sample sizes available, albeit some appraisals contain multiple comparisons for different endpoints. In most applications, weighting considerably reduces the effective sample size from the original $AC$ sample size. The median percentage reduction is 58\% (range: 7.9\% to 94.1\%; interquartile range: 42.2\% to 74.2\%). The final effective sample sizes are also representative of those in the technology appraisals, which are also small (median: 80; range: 4.8 to 639; interquartile range: 37 to 174).

\subsection{Estimands}\label{subsec43}


The estimand of interest is the marginal $A$ vs.~$B$ treatment effect in the $BC$ population. The treatment coefficient $\beta_T = \ln(0.25)$ is identical for both $A$ vs.~$C$ and $B$ vs.~$C$. Hence, the true conditional effect for $A$ vs.~$B$ in the $BC$ population is zero (subtracting the treatment coefficient for $A$ vs.~$C$ by that for $B$ vs.~$C$). Because the true unit-level treatment effects are zero for all subjects, the true marginal treatment effect in the $BC$ population is zero ($\Delta_{AB}^{*}=0$), which implies a ``null'' simulation setup in terms of the $A$ vs.~$B$ contrast, and average marginal and conditional effects for $A$ vs.~$B$ in the $BC$ population coincide by design. 

The simulation study meets the shared effect modifier assumption,\cite{phillippo2016nice} i.e., active treatments $A$ and $B$ have the same set of effect modifiers and the interaction effects $\beta_{2,k}$ of each effect modifier $k$ are identical for both treatments. Hence, the $A$ vs.~$B$ marginal treatment effect can be generalized to any given target population as effect modifiers are guaranteed to cancel out (the marginal effect for $A$ vs.~$B$ is conditionally constant across all populations). If the shared modifier assumption is not met, the true marginal treatment effect for $A$ vs.~$B$ in the $BC$ population will not be applicable in any target population (one has to assume that the target population is $BC$), and the average marginal and conditional effects for $A$ vs.~$B$ will likely not coincide as the measure of effect is non-collapsible.

\clearpage

\clearpage

\subsection{Methods}\label{subsec44}

Each simulated dataset is analyzed using the following methods:

\begin{itemize}
\item Matching-adjusted indirect comparison, as originally proposed by Signorovitch et al.,\cite{signorovitch2010comparative} where covariates are balanced for active treatment and control arms combined and weights are estimated using the method of moments. To avoid further reductions in effective sample size and precision, only the effect modifiers are balanced. A weighted Cox proportional hazards model is fitted to the IPD using the \texttt{R} package \texttt{survival}.\cite{therneau2000cox} Standard errors for the $A$ vs.~$C$ treatment effect are computed using a robust sandwich estimator\cite{signorovitch2010comparative, white1980heteroskedasticity} by setting \texttt{robust=TRUE} in \texttt{coxph}. Given the often arbitrary factors driving selection into different trials, the data-generating mechanism in subsection \ref{subsec42} does not specify a trial assignment model. Nevertheless, the logistic regression model for estimating the weights is considered approximately correct in that it selects the ``right'' subset of covariates as effect modifiers. The estimated weights are adequate for bias removal because the balancing property holds,\cite{dehejia1999causal, waernbaum2010propensity, zhao2008sensitivity, rubin2000combining} i.e., conditional on the weights, the effect modifier means are balanced between the two trials, and one can potentially achieve unbiased estimation of treatment effects in the $BC$ population.

\item Simulated treatment comparison:  a Cox proportional hazards regression on survival time is fitted to the IPD, with the IPD effect modifiers centered at the $BC$ mean values. The outcome regression is correctly specified. We include all of the covariates in the regression but only center the effect modifiers.  
\item The Bucher method\cite{bucher1997results} gives the standard indirect comparison. We know that this will be biased as it does not adjust for the bias induced by the imbalance in effect modifiers.  
\end{itemize}

In all methods, the variances of the within-trial relative effects are summed to estimate the variance of the $A$ vs.~$B$ treatment effect, $\hat{V}(\hat{\Delta}_{AB}^*)$. Confidence intervals are constructed using normal distributions: $\hat{\Delta}_{AB}^* \pm 1.96 \sqrt{\hat{V}(\hat{\Delta}_{AB}^*)}$, assuming relatively large $N$. 

\subsection{Performance measures}\label{subsec45}

We generate and analyze 1,000 Monte Carlo replicates of trial data per simulation scenario.  Let $\hat{\Delta}_{AB, s}^*$ denote the estimator for the $s$-th Monte Carlo replicate and let $\mathbb{E}(\hat{\Delta}_{AB}^*)$ denote its mean across the 1,000 simulations. Based on a test run of the method and simulation scenario with the highest long-run variability (MAIC under Scenario 109), we assume that $\textnormal{SD}(\hat{\Delta}_{AB}^*) \leq 0.45$ and that, conservatively, the variance across simulations of the estimated treatment effect is always less than approximately 0.2. Given that the Monte Carlo standard error (MCSE) of the bias is equal to $\sqrt{\textnormal{Var}(\hat{\Delta}_{AB}^*)/N_{sim}}$, where $N_{sim}$ is the number of simulations, it is at most 0.014 under 1,000 simulations. We consider the degree of precision provided by the MCSE to be acceptable in relation to the size of the effects. If the empirical coverage rate of the methods is 95\%, $N_{sim}=1000$ implies that the MCSE of the coverage is $\sqrt{(95 \times 5)/1000}=0.69\%$, with the worst-case MCSE being $1.58\%$ under 50\% coverage. We also consider this degree of precision to be acceptable. Hence, the simulation study is conducted under $N_{sim}=1000$.

The following criteria are considered jointly to assess the methods’ performances. MCSEs are estimated for each performance metric in order to quantify the simulation uncertainty due to using a finite number of simulation replicates.

\begin{itemize}
\item To assess aim 1, we compute the \textbf{bias} in the estimated treatment effect \[\mathbb{E}(\hat{\Delta}_{AB}^* - \Delta_{AB}^*) = \frac{1}{1000} \sum_{s=1}^{1000} \hat{\Delta}_{AB,s}^*-\Delta_{AB}^*.\] As $\Delta_{AB}^* = 0$, the bias is equal to the average estimated treatment effect across the simulations. The MCSE of the bias is estimated as $\sqrt{\frac{1}{1000 \times 999}  \sum_{s=1}^{1000}  (\hat{\Delta}_{AB,s}^*-\mathbb{E}(\hat{\Delta}_{AB}^*))}$.
\item To assess aim 2, we calculate the \textbf{variability ratio} of the treatment effect estimate, defined\cite{leyrat2014propensity} as the ratio of the average model standard error and the observed standard deviation of the treatment effect estimates (empirical standard error):
\begin{equation}
\mbox{VR}\left(\hat{\Delta}^*_{AB}\right)=\frac{\frac{1}{1000} \sum_{s=1}^{1000} \sqrt{\hat{V}(\hat{\Delta}^*_{AB,s})}}{\sqrt{\frac{1}{999}\sum_{s=1}^{1000} (\hat{\Delta}_{AB,s}^* - \mathbb{E}(\hat{\Delta}_{AB}^*))^2}}.
\label{eq11}
\end{equation}
$\mbox{VR}$ being greater than (or smaller) than one suggests that, on average, standard errors overestimate (or underestimate) the variability of the treatment effect estimate. It is important to note that this metric assumes that the correct estimand and corresponding variance are being targeted. A variability ratio of one is of little use if this is not the case, e.g.~if both the model standard errors and the empirical standard errors are taken over estimates targeting the wrong estimand. The MCSE of the variability ratio is approximated as:
\begin{equation*}
\sqrt{\frac{\frac{1}{1000}\sum_{j=1}^{1000} \Big (\sqrt{\hat{V}(\hat{\Delta}_{AB,j}^{\*})} - \mathbb{E}\big(\sqrt{\hat{V}(\hat{\Delta}_{AB}^{\*}) }\big)\Big )^2}{999 \times \mbox{ESE}(\hat{\Delta}^*_{AB})^2} + \frac{\Big(\frac{1}{1000} \sum_{j=1}^{1000} \sqrt{\hat{V}(\hat{\Delta}^{*}_{AB,j})}\Big)^2}{2 \times 999 \times \mbox{ESE}(\hat{\Delta}^*_{AB})^2}},
\end{equation*}
where $\mbox{ESE}(\hat{\Delta}^*_{AB})$ is the estimated empirical standard error, which is the denominator in Equation \ref{eq11}. 

\item Aim 3 is assessed using the \textbf{coverage} of confidence intervals, estimated as the proportion of times that the true treatment effect is enclosed in the $(100\times(1 - \alpha))\%$ confidence interval of the estimated treatment effect, where $\alpha=0.05$ is the nominal significance~level. The MCSE of the coverage is computed as $\sqrt{\frac{\mbox{Cover}(\hat{\Delta}^*_{AB}) \times (1 - \mbox{Cover}(\hat{\Delta}^*_{AB}))}{1000}}$, where $\mbox{Cover}(\hat{\Delta}^*_{AB})$ is the estimated coverage percentage.
\item We use \textbf{empirical standard error} (ESE) to assess aim 4 as it measures the precision or long-run variability of the treatment effect estimate. The ESE is defined above, as the denominator in Equation \ref{eq11}. The MCSE of the empirical standard error is estimated as $\frac{\mbox{ESE}(\hat{\Delta}^*_{AB})}{\sqrt{2 \times 999}}$.

\item The \textbf{mean square error} (MSE) of the estimated treatment effect \[\mbox{MSE}(\hat{\Delta}^*_{AB}) = \mathbb{E} \Big [ ( \hat{\Delta}^*_{AB} - \Delta^*_{AB})^2\Big ] = \frac{1}{1000} \sum_{s=1}^{1000}(\hat{\Delta}^*_{AB,s} - \Delta_{AB}^*)^2,\] provides a summary value of overall accuracy (efficiency), integrating elements of bias (aim 1) and variability (aim 4). The Monte Carlo standard error of the MSE is computed as $\sqrt{\frac{\sum_{s=1}^{1000}\Big[(\hat{\Delta}^*_{AB,s} - \Delta^*_{AB})^2 - \mbox{MSE}(\hat{\Delta}^*_{AB})\Big]^2}{1000 \times 999}}$, where $\mbox{MSE}(\hat{\Delta}^*_{AB})$ is the estimated mean square error.
\end{itemize}

\section{Results}\label{sec5}

The performance measures across all 162 simulation scenarios are illustrated in Figures \ref{fig2} to \ref{fig6} using nested loop plots,\cite{rucker2014presenting} which arrange all scenarios into a lexicographical order, looping through nested factors. In the nested sequence of loops, we consider first the parameters with the largest perceived influence on the performance metric. Notice that this order is considered on a case-by-case basis for each performance measure. Given the large number of simulation scenarios, depiction of Monte Carlo standard errors, quantifying the simulation uncertainty, is difficult. The Monte Carlo standard errors of each performance metric are reported in \hyperref[SC]{Supplementary Appendix C}. In MAIC, 1 of 162,000 weighted regressions had a separation issue, i.e., there is a total lack of covariate overlap (Scenario 115, with $N=150$). Results for this replicate were discarded. The outcome regressions converged for all replicates in STC and the Bucher method.

\subsection{Unbiasedness of treatment effect}\label{subsec51}

The impact of the bias will depend on the uncertainty in the estimated treatment effect,\cite{schafer2002missing, burton2006design} measured by the empirical standard error. To assess such impact, we consider standardizing the biases\cite{burton2006design} by computing these as a percentage of the empirical standard error. In a review of missing data methods, Schafer and Graham\cite{schafer2002missing} consider bias to be troublesome under 1,000 simulations if its absolute size is greater than about one half of the estimate’s empirical standard error, i.e., the standardized bias has magnitude greater than 50\%. Under this rule of thumb, MAIC does not produce problematic biases in any of the simulation scenarios. On the other hand, STC and the Bucher method generate problematic biases in 71 of 162 scenarios, and in 147 of 162 scenarios, respectively. The biases in MAIC do not appear to have any practical significance, as they do not degrade coverage and efficiency. 

\begin{figure}[!t]
\center{\includegraphics[width=0.93\textwidth]{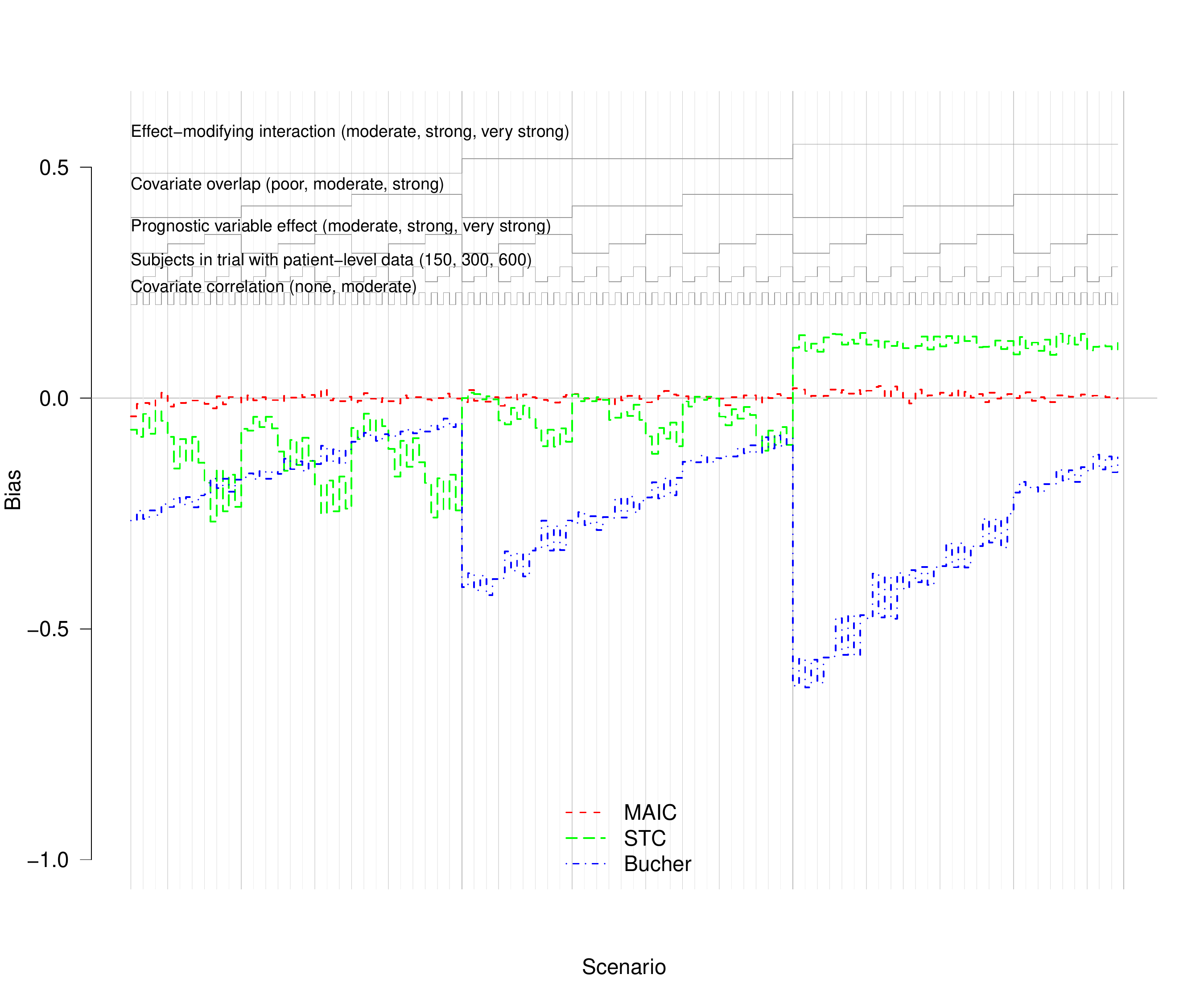}}
\caption{Bias across all simulation scenarios. The nested loop plot arranges all 162 scenarios into a lexicographical order, looping through nested factors. In the nested sequence of loops, we consider first the parameters with the largest perceived influence on the performance metric. MAIC: matching-adjusted indirect comparison; STC: simulated treatment comparison.\label{fig2}}
\end{figure}

Figure \ref{fig2} shows the bias for the methods across all scenarios. MAIC is the least biased method, followed by STC and the Bucher method. In the scenarios considered in this simulation study, STC produces negative bias when the interaction effects are moderate and positive bias when they are very strong. In addition, biases vary more widely when prognostic effects are larger. When interaction effects are weaker, stronger prognostic effects shift the bias negatively. This degree of systematic bias arises from the non-collapsibility of the (log) hazard ratio (see subsection \ref{subsec33}).

In some cases, e.g.~under very strong prognostic variable effects and moderate effect-modifying interactions, STC even has increased bias compared to the Bucher method. In other scenarios, e.g.~where there are strong effect-modifying interactions and moderate or strong prognostic variable effects, STC estimates are virtually unbiased. This is because, in these scenarios, the average conditional and marginal treatment effects for $A$ vs.~$C$ are almost identical and hence the non-collapsibility of the measure of effect is not an issue. It is worth noting that conclusions arising from the interpretation of patterns in Figure \ref{fig2} for STC are by-products of non-collapsibility. Any generalization should be cautious. 

As expected, the strength of interaction effects is an important driver of bias in the Bucher method and the incurred bias increases with greater covariate imbalance. This is because the more substantial the imbalance in effect modifiers and the greater their interaction with treatment, the larger the bias of the unadjusted comparison. The impact of these factors on the bias appears to be slightly reduced when prognostic effects are stronger and contribute more ``explanatory power'' to the outcome. Varying the number of patients in the $AC$ trial does not seem to have any discernible impact on the bias for any method. Biases in MAIC seem to be unaffected when varying the degree of covariate imbalance/overlap.

\subsection{Unbiasedness of variance of treatment effect}\label{subsec52}

\begin{figure}[!t]
\center{\includegraphics[width=\textwidth]{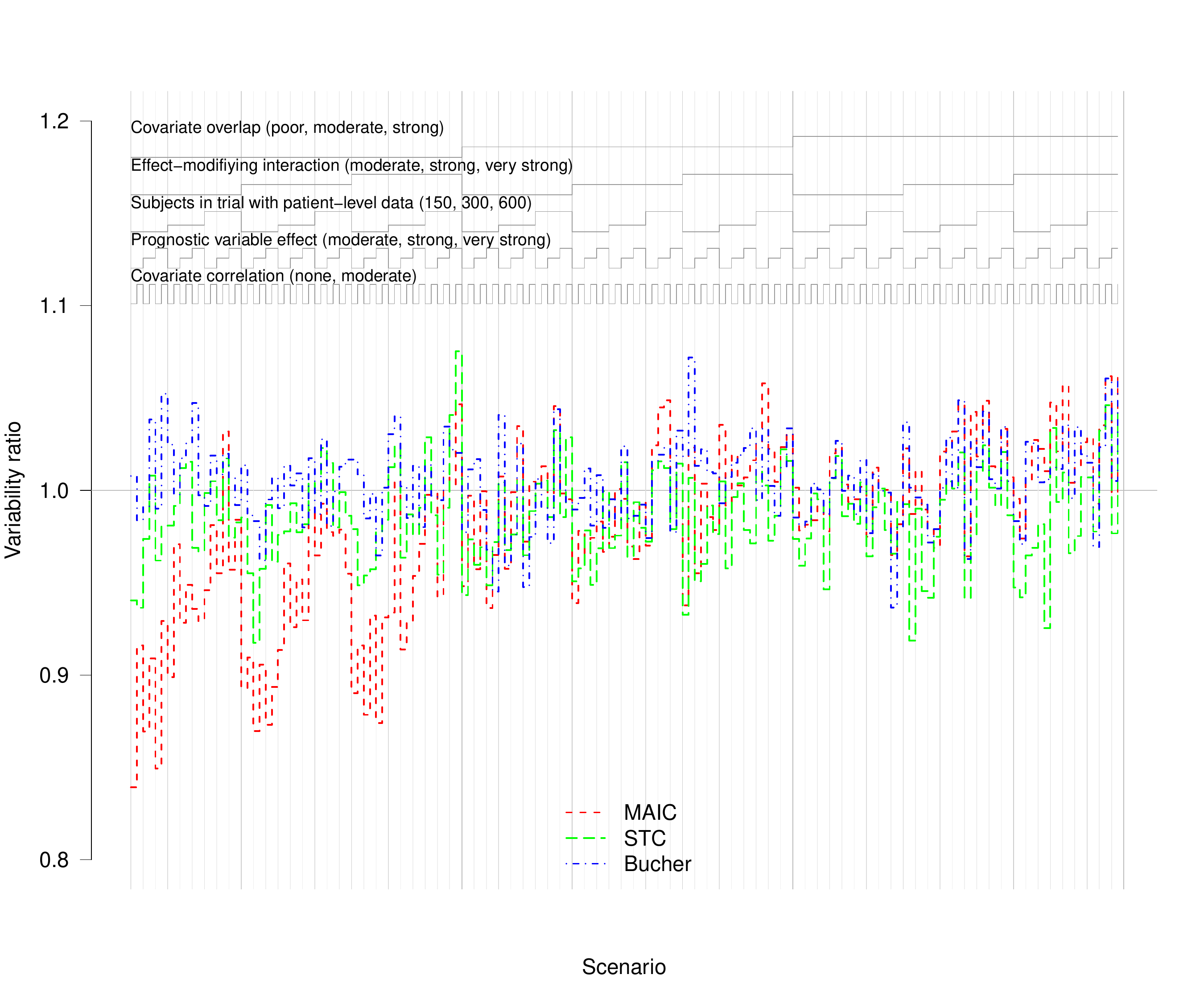}}
\caption{Variability ratio across all simulation scenarios. MAIC: matching-adjusted indirect comparison; STC: simulated treatment comparison.\label{fig3}}
\end{figure}

\begin{figure}[!b]
\center{\includegraphics[width=0.98\textwidth]{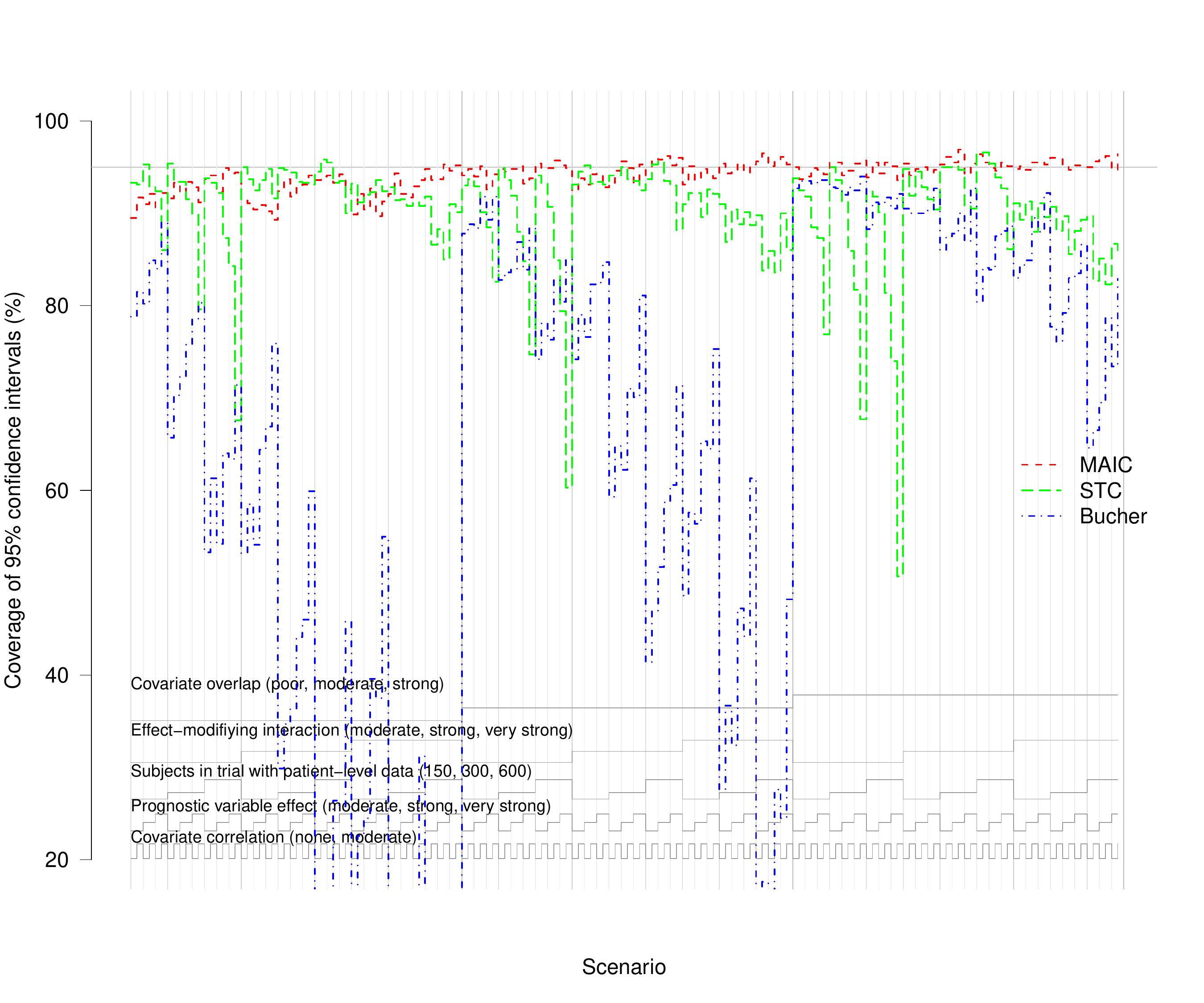}}
\caption{Empirical coverage percentage of 95\% confidence intervals across all simulation scenarios. MAIC: matching-adjusted indirect comparison; STC: simulated treatment comparison.\label{fig4}}
\end{figure}

In the Bucher method, the variability ratio is close to one under the vast majority of simulation scenarios (Figure \ref{fig3}). This suggests that standard error estimates for the methods are unbiased, i.e., that the model standard errors coincide with the empirical standard errors. In STC, variability ratios are generally close to one under $N=300$ and $N=600$, and any bias in the estimated variances appears to be negligible. However, the variability ratios decrease when the $AC$ sample size is small ($N=150$). In these scenarios, there is some underestimation of variability by the model standard errors. It is important to recall that this metric assumes that the correct estimand and corresponding variance are being targeted. This is not the case in our application of STC, in the sense that both model standard errors and empirical standard errors are taken over an incompatible indirect treatment comparison. MAIC standard errors underestimate variability when $N=150$, and also when covariate overlap is poor, in which case underestimation under $N=150$ is exacerbated. Under the smallest sample size and poor covariate overlap, variability ratios are often below 0.9, with model standard errors underestimating the empirical standard errors. This is likely due to the robust sandwich estimator used to derive the standard errors. In the literature, this has exhibited an underestimation of variability in small samples.\cite{kauermann2001note, fay2001small} The understated uncertainty is an issue, as it will be propagated through the cost-effectiveness analysis and may lead to inappropriate decision-making.\cite{claxton2005probabilistic} 

\subsection{Randomization validity}\label{subsec53}

From a frequentist viewpoint,\cite{neyman1934two} 95\% confidence intervals are randomization-valid if these are guaranteed to include the true treatment effect at least 95\% of the time. This means that the empirical coverage rate should be approximately equal to the nominal coverage rate, in this case 0.95 for 95\% confidence intervals, to obtain appropriate type I error rates for testing a ``no effect'' null hypothesis. Theoretically, the empirical coverage rate is statistically significantly different to 0.95 if, roughly, it is less than 0.9365 or more than 0.9635, assuming 1,000 independent simulations per scenario. These values differ by approximately two standard errors from the nominal coverage rate. When randomization validity cannot be attained, one would at least expect the interval estimates to be confidence-valid, i.e., the 95\% confidence intervals include the true treatment effect \textit{at least} 95\% of the time.

\clearpage

Confidence intervals from the Bucher method are not confidence-valid for virtually all scenarios. Coverage rates deteriorate markedly under the most important determinants of bias. When there is greater imbalance between the covariates and when interaction effects are stronger, the induced bias is larger and coverage rates are degraded. Under very strong interactions with treatment, empirical coverage may drop below 50\%. Therefore, the Bucher method will incorrectly detect significant results a large proportion of times in these scenarios. Such overconfidence will lead to very high type I error rates for testing a ``no effect'' null hypothesis.

In general, empirical coverage rates for MAIC do not overestimate the advertised nominal coverage rate. Only 4 of 162 scenarios have a rate above 0.9635. On the other hand, empirical coverage rates are significantly below the nominal coverage rate when the $AC$ sample size is low ($N=150$) and under poor covariate overlap. With $N=150$, 24 of 54 coverage rates are below 0.9365. When covariate overlap is poor, 38 of 54 coverage rates are below 0.9365 --- 18 of these under $N=150$. When there is both poor overlap and a low $AC$ sample size, coverage rates for MAIC are inappropriate: these may even fall below 90\%, i.e., at least double the nominal rate of error. Poor coverage rates are a decomposition of both the bias and the standard error used to compute the width of the confidence intervals. It is not bias that degrades the coverage rates for this method but the standard error underestimation mentioned in subsection \ref{subsec52}. Poor coverage is induced by the standard errors used in the construction of the confidence intervals. 

\subsection{Precision and efficiency}\label{subsec54}

\begin{figure}[!t]
\center{\includegraphics[width=\textwidth]{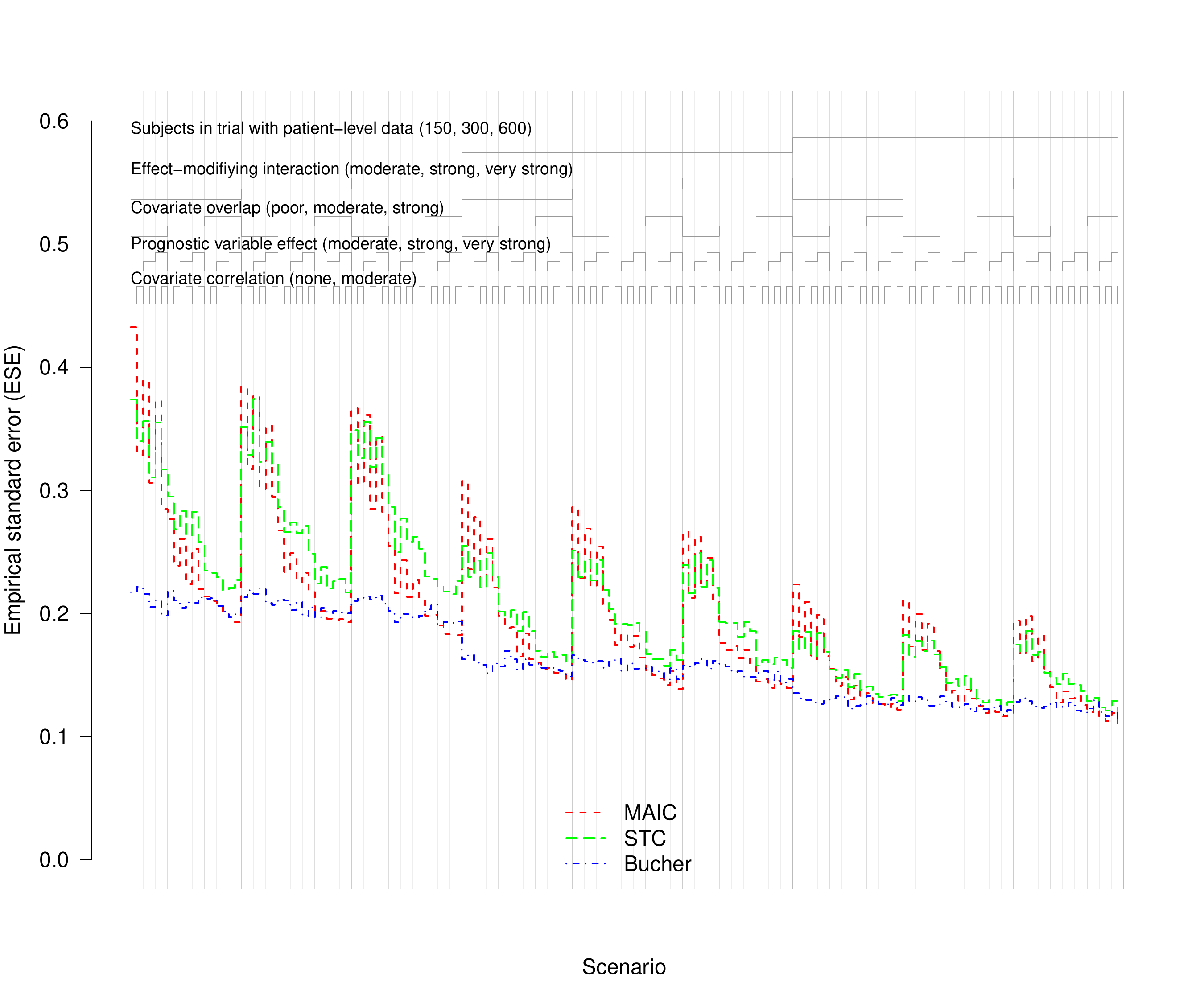}}
\caption{Empirical standard error across all simulation scenarios. MAIC: matching-adjusted indirect comparison; STC: simulated treatment comparison.\label{fig5}}
\end{figure}

Several trends are revealed upon visual inspection of the empirical standard error across scenarios (Figure \ref{fig5}). As expected, the ESE decreases for all methods (i.e., the estimate is more precise) as the number of subjects in the $AC$ trial increases. The strengths of interaction effects and of prognostic variable effects appear to have a negligible impact on the precision of population adjustment methods. 

\clearpage

The degree of covariate overlap has an important influence on the ESE and population adjustment methods incur losses of precision when covariate overlap is poor. When overlap is poor, there exists a subpopulation in $BC$ that does not overlap with the $AC$ population. Therefore, inferences in this subpopulation rely largely on extrapolation. Regression adjustment methods such as STC require greater extrapolation when the covariate overlap is poorer.\cite{phillippo2016nice} In reweighting methods such as MAIC, extrapolation is not even possible. When covariate overlap is poor, observations in the $AC$ patient-level data (those that are not covered by the range of the effect modifiers in the $BC$ population) are assigned very low weights (low odds of enrollment in $BC$ vs.~$AC$). On the other hand, the relatively small number of units in the overlapping region of the covariate space are assigned very large weights, dominating the reweighted sample. These extreme weights lead to large reductions in ESS and to the deterioration of precision and efficiency. 

In MAIC, the presence of correlation mitigates the effect of decreasing covariate overlap on a consistent basis. This is due to the correlation increasing the overlap between the joint covariate distributions of $AC$ and $BC$, lessening the reduction in effective sample size and providing greater stability to the estimates. ESE for the Bucher method does not vary across different degrees of covariate imbalance, as these are not considered by the method, and overprecise estimates are produced.

\begin{figure}[!b]
\center{\includegraphics[width=0.93\textwidth]{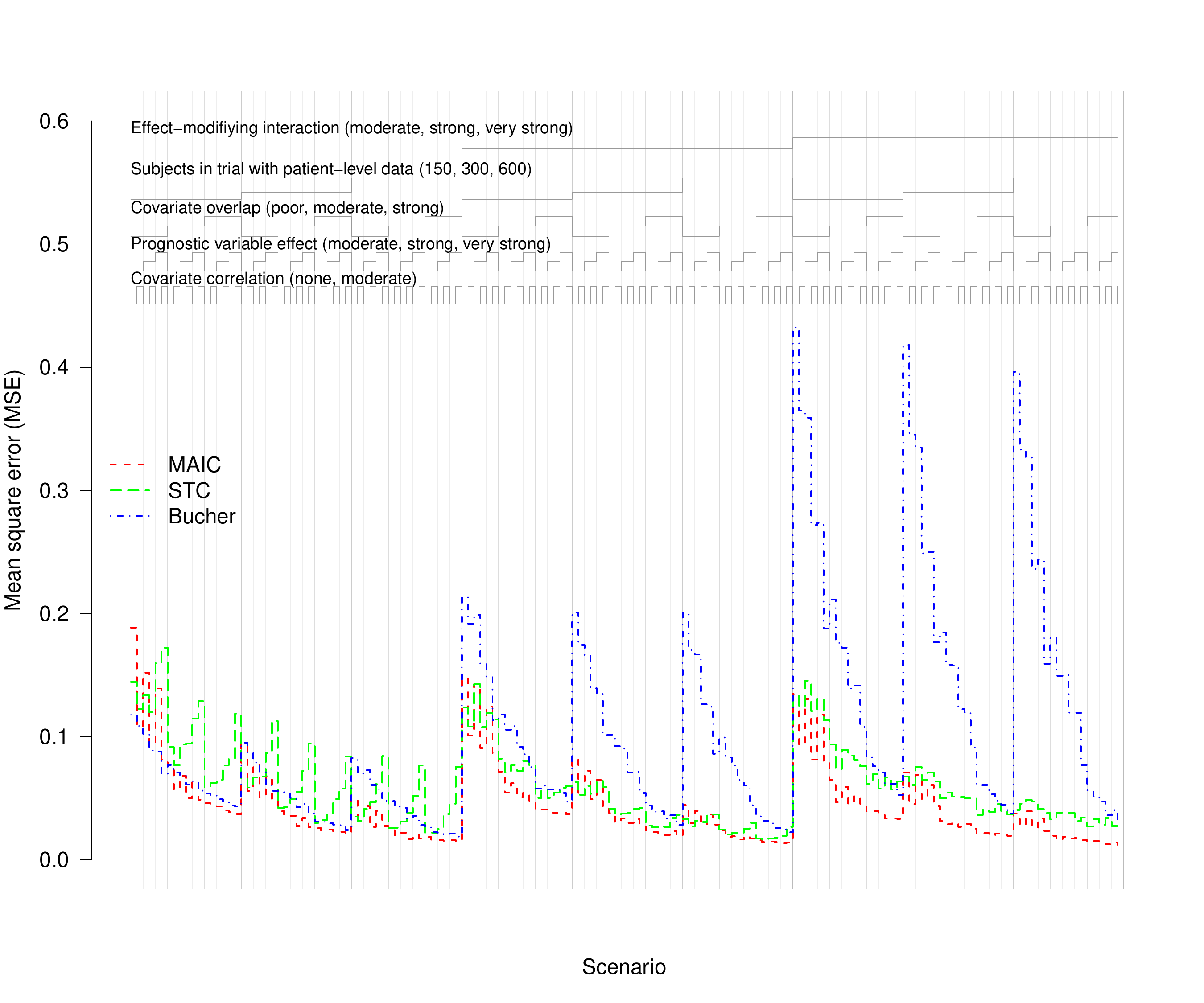}}
\caption{Mean square error across all simulation scenarios. MAIC: matching-adjusted indirect comparison; STC: simulated treatment comparison.\label{fig6}}
\end{figure}

Contrary to ESE, MSE also takes into account the true value of the estimand as it incorporates the bias. Hence, main drivers of bias and ESE are generally key properties for MSE. Figure \ref{fig6} is inspected in order to explore patterns in the mean square error. Estimates are less accurate for MAIC when prognostic variable effects are stronger, $AC$ sample sizes are smaller and covariate overlap is poorer. As bias is negligible for MAIC, precision is the driver of accuracy. On the contrary, as the Bucher method is systematically biased and overprecise, the driver of accuracy is bias. Poor accuracy in STC is also driven by bias, particularly under low sample sizes and strong prognostic variable effects. STC was consistently less accurate than MAIC, with larger mean square errors in all simulation scenarios. In some cases where the STC bias was strong, e.g.~very strong prognostic variable effects and moderate effect-modifying interactions, STC even increased the MSE compared to the Bucher method. 

\clearpage

In accordance with the trends observed for the ESE, the MSE is also very sensitive to the value of $N$ and decreases for all methods as $N$ increases. We highlight that the number of subjects in the $BC$ trial (not varied in this simulation study) is a less important performance driver than the number of subjects in $AC$; while it contributes to sampling variability, the reweighting or regressions are performed in the $AC$ patient-level data. 

\section{Discussion}\label{sec6}

In this section, we discuss the implications of, and recommendations for, performing population adjustment, based on the simulation study. Finally, we highlight potential limitations of the simulation study, primarily relating to the extrapolation of its results to practical guidance. We have seen in Section \ref{sec5} that the typical use of STC produces systematic bias as a result of the non-collapsibility of the log hazard ratio. The estimate $\hat{\Delta}_{AC}^*$ targets a conditional treatment effect that is incompatible with the estimate $\hat{\Delta}_{BC}$. This leads to bias in estimating the marginal treatment effect for $A$ vs $B$, despite all assumptions for population adjustment being met. Given the clear inadequacy of STC in this setting, we focus on MAIC as a population adjustment method.

An important future objective would be the development of an alternative formulation to STC that estimates a marginal treatment effect for $A$ vs.~$C$. A crucial additional step, missing from the current implementation, is to integrate or average the conditional effect estimates over the $BC$ covariates. Then, STC could potentially obtain a marginal treatment effect estimate that is comparable to the marginal $B$ vs.~$C$ estimate published in the $BC$ study. This would avoid the bias caused by incompatibility in the indirect comparison and provide inference for the marginal treatment effect for $A$ vs.~$B$ in the $BC$ population. During the preparation of this manuscript, a novel regression adjustment method named multilevel network meta-regression (ML-NMR) has been introduced.\cite{phillippo2020multilevel, phillippo2019calibration} ML-NMR targets a conditional treatment effect but directly avoids the compatibility issues of STC and is also applicable in treatment networks of any size with the two-study scenario as a special case. Using an averaging or integration step, ML-NMR could also be adapted to target a marginal treatment effect.\cite{phillippo2020target}   

\paragraph{Bias-variance trade-offs}

Before performing population adjustment, it is important to assess the magnitude of the bias induced by effect modifier imbalances. Such bias depends on the degree of covariate imbalance and on the strength of interaction effects, i.e., the effect modifier status of the covariates. The combination of these two factors determines the level of bias reduction that would be achieved with population adjustment. 

Inevitably, due to bias-variance trade-offs, the increase in variability that we are willing to accept with population adjustment depends on the magnitude of the bias that would be corrected. Such variability is largely driven by the degree of covariate overlap and by the $AC$ sample size. Hence, while the potential extent of bias correction increases with greater covariate imbalance, so does the potential imprecision of the treatment effect estimate (assuming that the imbalance is accompanied by poor overlap). 

In our simulation study, under no failures of assumptions, this trade-off always favors the bias correction offered by MAIC over the precision of the Bucher method, implying that the reductions in ESS based on unstable weights are worth it, even under stronger covariate overlap. Across scenarios, the relative accuracy of MAIC with respect to that of the Bucher method improves under greater degrees of covariate imbalance and poorer overlap. It is worth noting that, even in scenarios where the Bucher method is relatively accurate, it is still flawed in the context of decision-making due to overprecision and undercoverage. 

The magnitude of the bias that would be corrected with population adjustment also depends on the strength of interaction effects, i.e., the effect modifier status of the covariates. In the simulation study, the lowest effect-modifying interaction coefficient was $-\ln(0.67)=0.4$. Despite the relatively low magnitude of bias induced in this setting, MAIC was consistently more efficient than the Bucher method. Larger interaction effects warrant greater bias reduction but do not degrade the precision of the population-adjusted estimate. Hence, the relative accuracy of MAIC with respect to the Bucher method improves further as the effect-modifying coefficients increase. 

\paragraph{Justification of effect modifier status}

In the simulation study, we know that population adjustment is required as we set the cross-trial imbalances between covariates and have specified some of these as effect modifiers. Most applications of population adjustment present evidence of the former, e.g.~through tables of baseline characteristics with covariate means and proportions (``Table 1'' in a RCT publication). However, quantitative evidence justifying the effect modifier status of the selected covariates is rarely brought forward. Presenting this type of supporting evidence is very important when justifying the use of population adjustment. 

Typically, the selection of effect modifiers is supported by clinical expert opinion. However, clinical expert judgment and subject-matter knowledge are fallible when determining effect modifier status because: (1) the therapies being evaluated are often novel; and (2) effect modifier status is scale-specific --- clinical experts may not have the mathematical intuition to assess whether covariates are effect modifiers on the linear predictor scale (as opposed to the natural outcome scale). 

Therefore, applications of population adjustment often balance all available covariates on the grounds of expert opinion. This is probably because the clinical experts cannot rule out bias-inducing interactions with treatment for any of the baseline characteristics.  Almost invariably, the level of covariate overlap and precision will decrease as a larger number of covariates are accounted for in the analysis. Presenting quantitative evidence along with clinical expert opinion would help establish whether adjustment is necessary for each covariate.\cite{ricciardi2020dirichlet} 

As proposed by Phillippo et al.,\cite{phillippo2018methods} we encourage the analyst to fit regression models with interaction terms to the IPD for an exploratory assessment of effect modifier status. One possible strategy is to consider each potential effect modifier one-at-a-time by adding the corresponding interaction term to the main (treatment) effect model.\cite{harrell2016biostatistics} Then, the interaction coefficient can be multiplied by the difference in effect modifier means to gauge the level of induced bias.\cite{phillippo2016nice} This analysis should be purely exploratory, since individual trials are typically underpowered for interaction testing.\cite{fisher2011critical, fisher2017meta} The dichotomization or categorization of continuous variables, the poor representation of a variable, e.g.~a limited age range, and incorrectly assuming linearity may dilute interactions further.

Meta-analyses of multiple trials, involving the same outcome and similar treatments and conditions, provide greater power to detect interactions, particularly using IPD.\cite{fisher2017meta, tierney2015individual} With unavailable IPD, it may still be possible to conduct an IPD meta-analysis if the owners of the data are willing to provide the interaction effects,\cite{dias2018network} or one may conduct an ALD meta-analysis if covariate-treatment interactions are included in the clinical trial reports.\cite{fisher2011critical} In any case, the identification of effect modifiers is in essence observational,\cite{borenstein2011introduction, dias2011nice} and requires much more evidence than demonstrating a main treatment effect.\cite{remiro2020principled} Therefore, it may be reasonable to balance a variable if there is a strong biological rationale for effect modification, even if the interaction is statistically weak, e.g.~the $P$-value is large and the null hypothesis of interaction is not rejected.\cite{remiro2020principled} 

\paragraph{Nuances in the interpretation of results}

It is worth noting that the conclusions of this simulation study are dependent on the outcome and model type. We have considered survival outcomes and the Cox proportional hazards model, as these are the most prevalent outcome type and modeling framework in MAIC and STC applications. However, further simulation studies are required with alternative outcome types and models. For example, exploratory simulations with binary outcomes and logistic regression have found that the performance of MAIC is more affected by low sample sizes and poor covariate overlap than seen for survival outcomes. This is likely due to logistic regression being less efficient\cite{annesi1989efficiency} and more prone to small-sample bias\cite{vittinghoff2007relaxing} than Cox regression. 

Furthermore, we have only considered and adjusted for two effect modifiers that induce bias in the same direction, i.e., the effect modifiers in a given study have the same means, the cross-trial differences in means are in the same direction, and the interaction effects are in the same direction. In real applications of population adjustment, it is not uncommon to see more than 10 covariates being balanced.\cite{phillippo2019population} As this simulation study considered percentage reductions in effective sample size for MAIC that are representative of scenarios encountered in NICE TAs, real applications will likely have imbalances for each individual covariate that are smaller than those considered in this study. In addition, the means for the effect modifiers within a given study will differ, with the mean differences across studies and/or the effect-modifying interactions potentially being in opposite directions. Therefore, the induced biases could cancel out but, then again, this is not directly testable in a practical scenario. 

\paragraph{Potential failures in assumptions}

Most importantly, all the assumptions required for indirect treatment comparisons and valid population adjustment hold, by design, in the simulation study. While the simulation study provides proof-of-principle for the methods, it does not inform how robust these are to failures in assumptions. Population-adjusted analyses create additional complexity since they require a larger number of assumptions than standard indirect comparisons. The additional assumptions are hard to meet and most of them are not directly testable. It is important that researchers are aware of these, as their violation may lead to biased estimates of the treatment effect. In practice, we will never come across an idealistic scenario in which all assumptions perfectly hold. Therefore, researchers should exercise caution when interpreting the results of population-adjusted analyses. These should not be taken directly at face value, but only as tools to simplify a complex reality. 

Firstly, MAIC, STC and the Bucher method rely on trials $AC$ and $BC$ being internally valid, implying appropriate designs, the absence of non-compliance, proper randomization and reasonably large sample sizes.
Secondly, all indirect treatment comparisons (standard or population-adjusted) rely on consistency under parallel studies, i.e., potential outcomes are homogeneous for a given treatment regardless of the study assigned to a subject. For instance, treatment $C$ should be administered in the same setting in both trials, or differences in the nature of treatment should not change its effect. This means that MAIC and STC cannot account for cross-trial differences that are perfectly confounded with the nature of treatments, e.g.~treatment administration or dosing formulation. MAIC and STC can only account for differences in the characteristics of the trial populations. 

In practice, the additional assumptions made by MAIC and STC may be problematic. Firstly, it assumed that all effect modifiers for treatment $A$ vs.~$C$ are adjusted for.\footnote{In the anchored scenario, we are interested in a comparison of \textit{relative} outcomes or effects, not \textit{absolute} outcomes. Hence, an anchored comparison only requires conditioning on the effect modifiers, the covariates that explain the heterogeneity of the $A$ vs.~$C$ treatment effect. This assumption is denoted the \textit{conditional constancy of relative effects} by Phillippo et al.,\cite{phillippo2018methods, phillippo2016nice} i.e., given the selected effect-modifying covariates, the marginal $A$ vs.~$C$ treatment effect is constant across the $AC$ and $BC$ populations. There are analogous formulations of this assumption,\cite{stuart2011use, cole2010generalizing, kern2016assessing, pearl2014external, hartman2015sample} such as the conditional ignorability, unconfoundedness or exchangeability of trial assignment for such treatment effect, i.e., trial selection is conditionally independent of the treatment effect, given the selected effect modifiers. One can consider that being in population $AC$ or population $BC$ does not carry any information about the marginal $A$ vs $C$ treatment effect, once we condition on the treatment effect modifiers. This means that after adjusting for these effect modifiers, treatment effect heterogeneity and trial assignment are conditionally independent.} By design, the simulation study assumes that complete information is available for both trials and that all effect modifiers have been accounted for. In practice, this assumption is hard to meet --- it is difficult to ascertain the effect modifier status of covariates, particularly for new treatments with limited prior empirical evidence and clinical domain knowledge. Hence, the analyst may select the effect modifiers incorrectly. In addition, information on some effect modifiers could be unmeasured or unpublished for one of the trials. The incorrect omission of effect modifiers leads to the wrong specification of the trial assignment logistic regression model in MAIC, and of the outcome regression in STC. Relative effects will no longer be conditionally constant across trials and this will lead MAIC and STC to produce biased estimates. 

In the simulation study, we know the correct data-generating mechanism, and are aware of which covariates are purely prognostic variables and which covariates are effect modifiers. This is something that one cannot typically ascertain in practice. Exploratory simulations show that the relative precision and accuracy of MAIC deteriorate, with respect to STC and the Bucher method, if we treat all four covariates as effect modifiers. This is due to the loss of effective sample size and inflation of the standard error due to the overspecification of effect modifiers. 

Alternatively, it is more burdensome to specify the outcome regression model for STC than the propensity score model for MAIC; the outcome regression requires specifying both prognostic and interaction terms, while the trial assignment model in MAIC only requires the specification of effect modifiers. The relative precision and accuracy of STC deteriorate if the terms corresponding to the purely prognostic covariates are not included in the outcome regression. Nevertheless, this does not alter the conclusions of the simulation study: the other terms in the outcome regression already account for a considerable portion of the variability of the outcome and relative effects have very similar accuracy in any case.

Another assumption made by MAIC and STC, that holds in this simulation study, is that there is some overlap between the ranges of the selected covariates in $AC$ and $BC$. In population adjustment methods, the indirect comparison is performed in the $BC$ population. This implies that the ranges of the covariates in the $BC$ population should be covered by their respective ranges in the $AC$ trial. In practice, this assumption may break down if the inclusion/exclusion criteria of $AC$ and $BC$ are inconsistent. When there is no overlap, weighting methods like MAIC are unable to extrapolate beyond the $AC$ population, and may not even produce an estimate. However, STC can extrapolate beyond the covariate space observed in the $AC$ patient-level data, using the the linearity assumption or other appropriate assumptions about the input space. Note that the validity of the extrapolation depends on accurately capturing the true covariate-outcome relationships. We view extrapolation as a desirable property because poor overlap, with small effective sample sizes and large percentage reductions in effective sample size, is a pervasive issue in health technology appraisals.\cite{phillippo2019population}

MAIC and STC make certain assumptions about the joint distribution of covariates in $BC$. Where no correlation information is available for the $BC$ study, both methods seem to assume that the joint $BC$ covariate distribution is the product of the published marginal distributions. The implicit assumptions are, in fact, more nuanced. In MAIC, as stated in the NICE Decision Support Unit Technical Support Document,\cite{phillippo2016nice} ``when covariate correlations are not available from the ($BC$) population, and therefore cannot be balanced by inclusion in the weighting model, they are assumed to be equal to the correlations amongst covariates in the pseudo-population formed by weighting the ($AC$) population.'' In the typical usage of STC, the correlations between the $BC$ covariates are assumed to be equal to the correlations between covariates in the $AC$ study.  In the ``covariate simulation'' approach to STC, discussed in subsection \ref{subsec32}, this assumption is also made, albeit more explicitly, if the correlation structure observed in the $AC$ IPD is used to simulate the covariates. In an anchored comparison, only effect-modifying covariates need balancing, so the assumptions can be relaxed to only include effect modifiers. This set of assumptions will only induce bias if higher-order interactions (involving two or more covariates) are unaccounted for or misspecified. If these interactions are not included in the weighting model for MAIC or in the outcome regression for STC, the specification of pairwise correlations will not make a difference in terms of bias, as observed in a recent simulation study that investigates this set of assumptions.\cite{phillippo2020multilevel} 

All indirect treatment comparisons should be performed and are typically conducted on the linear predictor scale,\cite{phillippo2016nice} upon which the effect of treatment is assumed to be additive. We have assumed that the effect modifiers have been defined on the linear predictor scale and are additive on this scale. In the simulation study, it is known that effect modification is linear on the log hazard ratio scale. A central component of population-adjusted indirect comparisons is the specification of a model that is typically parametric. That is the propensity score model for the weights in MAIC or the outcome regression in STC. Parametric modeling assumptions may not be appropriate in real applications, where there is a danger of model misspecification. This is more evident in a regression adjustment method like STC, where an explicit outcome regression is formulated. The parametric model depends on functional form assumptions that will be violated if the covariate-outcome relationships are not correctly captured.

Even though the logistic regression model for the weights in MAIC does not make reference to the outcome, MAIC is also susceptible to model misspecification bias, albeit in a more implicit form. The model for estimating the weights is approximately correct in the simulation study because the right subset of covariates has been selected as effect modifiers and the balancing property holds for the weights, as mentioned in subsection \ref{subsec44}. In practice, the model will be incorrectly specified if this is not the case, potentially leading to a biased estimate. Scale conflicts may also arise if effect modification status, which is scale-specific, has been justified on the wrong scale, e.g.~when treatment effect modification is specified as linear but is non-linear or multiplicative, e.g.~age in cardiovascular disease treatments. Note that, in practice, we find that it may be more difficult to specify a correct parametric model for the outcome than an approximately correct parametric model for the trial assignment weights. 

Finally, population-adjusted indirect comparisons only produce an estimate $\hat{\Delta}_{AB}^*$ that is valid in the $BC$ population, which may not match the target population for the decision unless an additional assumption is made. This is the shared effect modifier assumption,\cite{phillippo2016nice} described in subsection \ref{subsec43}. This assumption is met by the simulation study and is required to transport the treatment effect estimate to any given target population. However, it is untestable for MAIC and STC with the data available in practice. Shared effect modification is hard to meet if the competing interventions do not belong to the same class, and have dissimilar mechanisms of action or clinical properties. In that case, there is little reason to believe that treatments $A$ and $B$ have the same set of effect-modifying covariates and that these interact with active treatment in the same way in $AC$ and $BC$. It is worth noting that the target population may not match the $AC$ and $BC$ trial populations and may be more akin to the joint covariate distribution observed in a registry/cohort study or some other observational dataset. Policy-makers could use such data to define a target population for a specific outcome and disease area into which all manufacturers could conduct their indirect comparisons. This would help relax the shared effect modifier assumption. 

Given the large number of assumptions made by population-adjusted indirect comparisons, future simulation studies should assess the robustness of the methods to failures in assumptions under different degrees of data availability and model misspecification.

\paragraph{Variance estimation in MAIC}

MAIC was generally randomization-valid, except in situations with poor covariate overlap and small sample sizes, where robust sandwich standard errors underestimated empirical estimates of the standard error and, consequently, there was undercoverage. MAIC exhibited variability ratios below 0.9 in scenarios with the smallest sample size and poor covariate overlap. In these scenarios, confidence intervals were narrow, achieving coverage rates which were statistically significantly below 95\% and sometimes dropping below 90\%. As mentioned in subsection \ref{subsec52}, this is probably due to the robust sandwich estimator used to derive the standard errors, which has previously underestimated variability in small samples in simulation studies.\cite{kauermann2001note, fay2001small} It is worth noting that this estimator is based on large-sample (asymptotic) arguments and infinite populations. Therefore, it is not surprising that performance is poor under the smallest effective sample sizes, which occur where the $AC$ trial sample size is small and covariate overlap is poor. Where effective sample sizes are small, confidence intervals derived from robust sandwich variance estimators should be interpreted cautiously, as these may understate uncertainty and this underestimation will be propagated through the cost-effectiveness analysis, potentially leading to inappropriate decision-making.

This robust variance estimator is easy to use by analysts performing MAIC (and propensity score weighting, in general) because it is computationally efficient and is typically implemented in standard routines in statistical computing software such as \texttt{R}. For instance, in \texttt{R}, by setting \texttt{robust=TRUE} in the \texttt{coxph} function, built in the \texttt{survival} package\cite{therneau2000cox} for survival analysis, or using the \texttt{sandwich} package\cite{zeileis2004econometric} for the treatment coefficient of a weighted generalized linear model. It is worth noting that these readily available implementations assume that the weights are fixed or known and do not account for the uncertainty in the estimation of the weights. 

In principle, one could circumvent this issue by using the bootstrap to obtain the variance and confidence intervals of the $A$ vs.~$C$ treatment effect, as in the simulation study by Petto et al.\cite{petto2019alternative} or in the article by Sikirica et al.\cite{sikirica2013comparative} Bootstrap methods are beneficial because they can account for the variability of the estimated weights and are straightforward to implement, potentially providing unbiased variance estimators with a large number of resamples. However, bootstrapping is orders of magnitude more expensive computationally than applying the closed-form sandwich variance estimator. In addition, bootstrap resampling procedures are inherently random and exhibit some seed-dependence, which is only mitigated by increasing the number of resamples and computational demand. Future simulation studies should compare different approaches to variance estimation and assess whether implementations of the bootstrap can compete with the robust sandwich estimator.

Another alternative to variance estimation is the development of closed-form robust sandwich estimators that properly account for the uncertainty in estimating the propensity score logistic regression for the weights. These have been explicitly derived for accurate variance estimation in the causal inference literature,\cite{lunceford2004stratification, buchanan2018generalizing, li2019propensity, mao2019propensity} but not for MAIC. This is a priority for future research. 

\paragraph{Unanchored comparisons}

Finally, it is worth noting that, while this article focuses on anchored indirect comparisons, most applications of population adjustment in HTA are in the unanchored setting,\cite{phillippo2019population} both in published studies and in health technology appraisals. We stress that RCTs deliver the gold standard for evidence on efficacy and that unanchored comparisons make very strong assumptions which are largely considered impossible to meet (absolute effects are conditionally constant as opposed to relative effects being conditionally constant).\cite{phillippo2018methods, phillippo2016nice} Unanchored comparisons effectively assume that absolute outcomes can be predicted from the covariates, which requires accounting for all variables that are prognostic of outcome. 

However, the number of unanchored comparisons is likely to continue growing as regulators such as the United States Food and Drug Administration and the European Medicines Agency are, increasingly, and particularly in oncology, approving new treatments on the basis of observational or single-armed evidence, or disconnected networks with no common comparator.\cite{hatswell2016regulatory, beaver201825} As pharmaceutical companies use this type of evidence to an increasing extent to obtain accelerated or conditional regulatory approval, reimbursement agencies will, in turn, be increasingly asked to evaluate interventions where only this type of evidence is available. Therefore, further examinations of the performance of population adjustment methods must be performed in the unanchored setting. 

\paragraph{Areas of debate}

Population-adjusted indirect comparisons make up a major area of methodological developments in evidence synthesis, with applications in HTA worldwide. We acknowledge that there is still debate in some areas, which may require further study. It is claimed that, for the (log) hazard ratio, marginal treatment effects may vary across different distributions of the purely prognostic covariates. Hence, these covariates can still modify marginal treatment effects, even in the absence of interaction effects, and cross-trial differences in these can potentially induce bias. In our simulation study, MAIC does not account for imbalances in purely prognostic variables (these are the covariates with only main effects, not interaction effects, in the Cox model) and remains unbiased. Nevertheless, this remains a topic for further investigation. 

Another area of debate is whether marginal or conditional effects are more appropriate target estimands for population-level decision-making in HTA.\cite{phillippo2020target, remiro2020conflating, phillippo2020assessing} We endorse the use of marginal effects as population-level estimates, required for reimbursement decisions at the population level.\cite{remiro2020conflating} Nevertheless, conditional treatment effect estimates, adjusted for prognostic factors, have been termed ``population-average'' effects,\cite{phillippo2020target, phillippo2020assessing} and recommended on the grounds of: (1) providing more statistically precise and efficient decision-making; and (2) the clinical trials literature preferring ``covariate-adjusted'' over ``unadjusted'' analyses in order to account for the distribution of covariates.\cite{phillippo2020target} We note that these conclusions are based on covariate adjustment using linear regression and continuous outcomes, and on conflating the terms ``marginal'' and ``unadjusted''. Firstly, when working with non-collapsible effect measures such as odds ratios in logistic regression with binary outcomes, or hazard ratios in Cox regression with survival outcomes, conditional covariate-adjusted treatment effect estimates actually reduce precision and efficiency with respect to unadjusted marginal estimates, in the ``ideal RCT'' analysis.\cite{daniel2020making, robinson1991some, ford1995model, hernandez2004covariate} Secondly, it is worth noting that marginal need not mean unadjusted.\cite{daniel2020making} Marginal effects can also be covariate-adjusted and, in fact, population-adjusted indirect comparisons should ultimately produce covariate-adjusted marginal effect estimates, that account for the relevant covariate distribution and, for non-collapsible effect measures, can potentially increase precision and efficiency with respect to both conditional and unadjusted marginal effect estimates.\cite{daniel2020making, tsiatis2008covariate, benkeser2020improving, colantuoni2015leveraging, diaz2016enhanced}

\section{Concluding remarks}\label{sec7}

In the performance measures we considered, MAIC was the least biased and most accurate method under no failures of assumptions. We therefore recommend its use for survival outcomes, provided that its assumptions are reasonable. MAIC was generally randomization-valid, except in situations with poor covariate overlap and small sample sizes (small effective sample sizes), where robust sandwich standard errors underestimated variability and there was undercoverage.  

The typical usage of STC produced systematic bias because it targeted a conditional treatment effect for $A$ vs.~$C$, where the target estimand should be different, a marginal treatment effect. The conditional treatment effect is incompatible in the indirect comparison. Note that STC is not intrinsically biased; it simply targets the wrong estimand in this setting. If we intend to target a marginal treatment effect for $A$ vs.~$C$ and naively assume that this version of STC does so, there will be bias due to the non-collapsibility of the log hazard ratio. This bias could have considerable impact on decision making and policy, and could lead to perverse decisions and subsequent misuse of resources. Therefore, the typical use of STC should be avoided, particularly in settings with a non-collapsible measure of effect. An important future objective would be the development of an alternative formulation to STC that estimates a marginal treatment effect for $A$ vs.~$C$. A crucial additional step, missing from the current implementation, is to integrate or average the conditional effect estimates over the $BC$ covariates. Then, STC could potentially obtain a marginal treatment effect estimate that is compatible with the marginal $B$ vs.~$C$ estimate published in the $BC$ study. 

The Bucher method is systematically biased and overprecise when there are imbalances in effect modifiers and interaction effects that induce bias in the treatment effect. Future simulation studies should evaluate population adjustment methods with different outcome types and when assumptions fail. 

\section*{Acknowledgments}

The authors thank Anthony Hatswell for discussions that contributed to the quality of the manuscript and acknowledge Andreas Karabis for his advice and expertise in MAIC. Part of these findings are to be presented as a research article in a peer-reviewed journal. The authors thank the editor and peer reviewers of the article throughout this process. Their comments were hugely insightful and substantially improved the article, for which the authors are grateful. Finally, the authors thank Tim Morris, who provided very helpful comments after evaluating Antonio Remiro Az\'ocar's PhD proposal defense. This article is based on research supported by Antonio Remiro-Az\'ocar's PhD scholarship from the Engineering and Physical Sciences Research Council of the United Kingdom. Gianluca Baio was partially funded by a research grant sponsored by Mapi/ICON at University College London. Anna Heath was funded through an Innovative Clinical Trials Multi-year Grant from the Canadian Institutes of Health Research (funding reference number MYG-151207; 2017 - 2020).

\subsection*{Financial disclosure}

Funding agreements ensure the authors’ independence in designing the simulation study, interpreting the results, writing, and publishing the article.

\subsection*{Conflict of interest}

The authors declare no potential conflict of interests.

\subsection*{Data Availability Statement}

The files required to generate the data, run the simulations, and reproduce the results are available at \url{http://github.com/remiroazocar/population_adjustment_simstudy}.

\clearpage

\section*{Supplementary Appendix A: Method assumptions}\label{SA}
\addcontentsline{toc}{section}{Supplementary Appendix A: Method assumptions}

In practice, we consider that a causal treatment effect is of interest in any indirect comparison between treatments, which ultimately seeks to mimic the analysis that would be conducted in a head-to-head RCT. This discussion of assumptions borrows terminology from the potential outcomes framework for causal inference, commonly known as the Rubin Causal Model,\cite{holland1986statistics} originally proposed by Neyman\cite{neyman1923application} in the analysis of randomized experiments, and generalized to observational studies by Rubin.\cite{rubin1974estimating} We introduce the assumptions made by population-adjusted indirect comparisons in this context. Note that this discussion is non-technical and detailed theory and notation based on potential outcomes are not presented. 

Population-adjusted indirect comparisons make the following assumptions, required to make valid causal inferences in the $BC$ population: (1) internal validity; (2) consistency under parallel studies; (3) conditional strong ignorability of trial assignment for the $A$ vs.~$C$ treatment effect (this requires both the conditional constancy of relative effects and overlap/positivity across the covariate distributions); (4) correct specification of the $BC$ population; and (5) linear (parametric) modeling assumptions. The first two assumptions are shared by MAIC, STC and the Bucher method; in fact, these are made by any indirect comparison or meta-analysis. Assumptions that are not specific to indirect treatment comparisons, e.g.~those that are specific to the type of regression model used, such as proportional hazards or non-informative censoring for a Cox regression, are not discussed.  

Note that the following assumptions can only guarantee an unbiased comparison if the within-trial relative effects target compatible estimands of the same type. The majority of RCTs publish an estimate for $B$ vs.~$C$ that targets a marginal treatment effect (any published estimate of a conditional treatment effect is likely incompatible with that for $A$ vs.~$C$). Hence, $\hat{\Delta}_{AC}$ (for the Bucher method) or $\hat{\Delta}^*_{AC}$ (for the population adjustment methods) should target a marginal treatment effect. If a comparison of conditional treatment effects is performed, these would have to be adjusted across identical sets of covariates, using the same model specification. 


Those studying the generalizability of treatment effects often make a distinction between sample-average and population-average marginal effects.\cite{stuart2011use, cole2010generalizing, kern2016assessing, hartman2015sample} Typically, another implicit assumption made by population-adjusted indirect comparisons is that the marginal treatment effects estimated in the $BC$ sample, as described by its published covariate moments in the case of the $A$ vs.~$C$ treatment effect, coincide with those that would be estimated in the target population of the trial. Namely, either the study sample on which inferences are made is the study target population, or it is a simple random sample (i.e., representative) of such population, ignoring sampling variability. 

In addition, it is worth noting that when referring to ``effect modifiers'', we refer to the variables that modify the treatment effect measure for $A$ vs.~$C$ in the linear predictor scale. We select the effect modifiers of treatment $A$ with respect to $C$ (as opposed to the effect modifiers of treatment $B$ with respect to $C$), because we have to adjust for these to perform the indirect comparison in the $BC$ population, implicitly assumed to be the target population. If we had IPD for the $BC$ study and ALD for the $AC$ study, we would have to account for the covariates that modify the effect of treatment $B$ vs.~$C$, in order to perform the comparison in the $AC$ population.

\subsection*{Internal validity}

The first set of assumptions relates to the internal validity of the $AC$ and $BC$ trials. The trials are internally valid under the following structural assumptions, which are necessary for causal inference:

\begin{itemize}
\item Stable unit treatment value assignment (SUTVA). This assumption implies that: (1) the treatment of a given subject does not affect the potential outcomes of other individuals (non-interference);\cite{rubin1980randomization, hudgens2008toward} and (2) there is only one version of each treatment (treatment-variation irrelevance),\cite{vanderweele2009concerning} implying that the treatment is comparable across units.\cite{vanderweele2013causal} The first condition is questionable, for example, in a vaccine trial, where the outcome of an individual (i.e., developing the flu) depends on the vaccination status of others because of herd immunity. The second condition is questionable if there are differences among versions of treatment, e.g.~in the delivery mechanism, that are relevant to the outcome of interest.
\item Strongly ignorable treatment assignment. Ignorability implies that treatment assignment is independent of the potential outcomes.\cite{rosenbaum1983central} Ignorability can be conditional on the observed baseline characteristics or unconditional. Conditional ignorability is strong when there is positivity\cite{hernan2006estimating} or overlap, i.e., any subject has a positive probability of being assigned to either treatment group given the baseline covariates. 
\end{itemize}

The SUTVA assumption is met by appropriate study design.\cite{rubin2005causal} By design, the conditions of positivity\cite{cole2008constructing} and ignorability,\cite{greenland2009identifiability} whether this is plain or conditional on baseline covariates, are met by randomized trials. The random allocation of treatment ensures that, on expectation, there are no systematic differences in the distribution of (measured and unmeasured) baseline covariates between treatment groups, i.e., there is covariate balance.\cite{austin2013performance, greenland1990randomization} Note that balance is a large sample property. In small samples, one may still observe residual differences in baseline characteristics that bias the comparison. 
Therefore, the internal validity assumptions are met if the $AC$ and $BC$ trials are appropriately designed studies with proper randomization and reasonably large sample sizes. Internal validity is sufficient for unbiased estimation of $\Delta_{AC}$ and $\Delta_{BC}$ within each trial. Finally, we have assumed that internal validity in each trial is not compromised by other issues, such that there is negligible measurement error or missing data, perfect adherence to treatment, the absence of non-compliance, etc.

In the unanchored case with single-arm studies, strongly ignorable treatment assignment is not required as there is no common comparator arm. However, unanchored comparisons are subject to the additional assumptions and biases of non-randomized study designs, which are often stronger.\cite{deeks2003evaluating}

\subsection*{Consistency under parallel studies}

Consistency under parallel studies\cite{hartman2015sample} is the cross-trial version of the second condition of SUTVA (treatment-variation irrelevance). This assumption implies that potential outcomes for an individual under a given treatment are homogeneous regardless of the study assigned to the individual. For instance, treatment $C$ should be administered in the same setting in both trials, or differences in the nature of treatment, e.g.~in the clinical protocol or delivery mechanism, should not change its effect. In there are non-negligible differences in the versions of treatment, for instance, if treatment $C$ is accompanied by adherence counseling in one of the trials, while such counseling is absent in the other, this assumption could be invalid. 

Consistency under parallel studies means that MAIC and STC cannot adjust for cross-trial differences related to the nature of treatments, e.g.~treatment administration, switching, dosing formulation, titration or co-treatments. Differences of this type are perfectly confounded with treatment\cite{phillippo2016nice} and MAIC and STC can only adjust for differences in trial population characteristics. This assumption is required to perform any valid indirect comparison across studies. 
\subsection*{Conditionally strong ignorability of trial assignment}

Strongly ignorable trial assignment (specifically, assignment to the $AC$ trial), is the primary assumption underlying population-adjusted indirect comparisons and is required for unbiased estimation of $\Delta_{AC}^*$. This is akin to the strongly ignorable sample assignment assumption\cite{hartman2015sample} commonly used in the generalizability literature.\cite{hartman2015sample, cole2010generalizing, stuart2011use, kern2016assessing} Generalizability seeks to extend relative treatment effects obtained from a RCT into a target population --- the trial sample is considered to be an unrepresentative subpopulation of the, more diverse, target population. In MAIC and STC, the indirect comparison is performed in the $BC$ population, and the $A$ vs.~$C$ treatment effect is ``generalized'' to the $BC$ population. Strong ignorability collectively relies on two assumptions: ignorability and overlap (or positivity). Note that, even though strong ignorability has been proposed in the context of propensity score modeling, it is also required for robust causal inference in the $BC$ population when using regression adjustment approaches to estimate treatment effects.

\subsubsection*{Conditional ignorability}

This assumption has many possible formulations. In the anchored case, one can consider that trial assignment/selection is conditionally ignorable, unconfounded or exchangeable for the $A$ vs.~$C$ treatment effect (the potential relative outcomes for $A$ vs.~$C$), i.e., being in study $AC$ or study $BC$ does not carry over any information about such treatment effect, once we condition on the selected treatment effect modifiers. This means that after accounting for these effect modifiers, treatment effect heterogeneity and trial assignment are conditionally independent. 

MAIC will only meet conditional ignorability if \textit{all} (observed or unobserved) effect modifiers are accounted for, regardless of whether these are balanced before the weighting (excluding balanced covariates from the weighting procedure does not ensure balance after the weighting). STC meets conditional ignorability if all \textit{imbalanced} effect modifiers are accounted for in the regression model. 

This assumption is challenging to meet in practice and also untestable. On one hand, it is tied to the scale used to define the treatment effects and effect modifiers. Most importantly, it requires that all effect modifiers are measured in the IPD for study $AC$ and that the set of published baseline characteristics for study $BC$ is sufficiently rich, such that all effect modifiers are available. This is a key challenge as ``Table 1'' data on the published RCT for the $BC$ population are often limited, and it can sometimes be difficult to find common measures between different studies. In addition, all effect modifiers must be accounted for by the analyst, who may select the effect modifiers incorrectly. It is generally difficult to ascertain the effect modifier status of variables, particularly for new treatments with limited prior empirical evidence and clinical domain knowledge. Overspecification of effect modifiers will not bias the comparison but may inflate standard errors and lead to a subsequent loss of precision.  

In the unanchored case, there is no common comparator group included in the analysis. Therefore, estimates are based on a comparison of within-trial absolute outcomes from single treatment arms, obtained from single-arm studies or individual arms of observational studies or RCTs, not on a comparison of within-trial relative effects. In this scenario, trial assignment is ignorable if it is conditionally independent of the potential absolute outcomes given the covariates accounted for in the adjustment mechanism. 

Here, we cannot draw a distinction between the predictors of outcome that are not treatment-specific (prognostic variables), i.e., associated with outcomes on follow-up regardless of the treatment provided, and factors that are associated with the outcome under a specific intervention because of interaction with treatment (effect modifiers). In fact, treatment effect modification cannot be quantified and is ill-defined. This is because its definition is reliant on contrasting outcomes between two groups, and there is no reference control group to define a relative treatment effect in the IPD trial.\footnote{If the IPD study is a single-arm trial, it is not possible to determine whether the outcomes are due to strong prognostic indicators or to the treatment itself and its interaction with effect modifiers, in excess of prognostic impact. Treatment effect modifiers cannot be identified because single-arm trials do not provide information about outcomes in a control group not receiving the intervention. Even if the index trial is an RCT comparing treatments $A$ and $C$, effect modification for the relative effect of $A$ vs $C$ does not translate across studies if there is not a common comparator group. For instance, if the comparator study contrasts treatments $B$ and control $D$, a covariate that modifies the effect of active intervention $A$ with respect to $C$ is not necessarily an effect modifier with respect to $D$.} 
The effect of purely prognostic variables and variables that are predictive of response to a specific treatment is conflated and cannot be disentangled. Therefore, unanchored comparisons rely on the assumption of no systematic cross-trial differences in predictors of the absolute outcome under treatment $A$, regardless of whether they have a purely prognostic role or are predictors of the response to treatment. The population adjustment methods only meet ignorability in the unanchored case if all variables that are prognostic of outcome under treatment $A$ are balanced. In the unanchored case, ignorable trial assignment is equivalent to the conditional constancy of absolute effects described in the literature.\cite{phillippo2016nice, phillippo2018methods}

\subsubsection*{Overlap}

Conditional ignorability of trial assignment is strong if there is positivity or overlap, i.e., if every subject in the $BC$ population has a positive probability of being assigned to the $AC$ trial given the covariates accounted for in the adjustment mechanism. This implies that the ranges of the covariates in the $BC$ population are covered by their respective ranges in the $AC$ trial. This assumption may pose a problem if the inclusion/exclusion criteria of $AC$ and $BC$ are inconsistent. For instance, consider a situation where age is selected as an effect modifier and the age ranges of trial $AC$ and trial $BC$ are 60-70 and 40-70, respectively. There exists a subpopulation (age 40-60) in $BC$ that does not overlap with the $AC$ population. Hence, the $AC$ study provides no evidence or information about the treatment effect and treatment effect modification in the excluded age group, and $\hat{\Delta}_{AC}^*$ may be biased in the full comparator trial population (ages 40-70). 

In such cases, weighting methods like MAIC are unable to extrapolate beyond the observed covariate space in the $AC$ IPD, as there are no subjects to reweight. Where overlap is insufficient, a regression adjustment method like STC can extrapolate beyond the $AC$ population, using the linearity assumption or other appropriate assumptions about the input space. However, valid extrapolation requires accurately capturing the true relationship between the covariates and the outcome.

Conversely, the exclusion of patients enrolled in $AC$ from the $BC$ population, e.g.~if the $AC$ population is more diverse, does not necessarily violate the overlap assumption in MAIC and STC. This is because these methods deliver estimates in the $BC$ population. Therefore, adjustment in this scenario is an interpolation as opposed to an extrapolation of the observed subject-level data for $AC$. In this scenario, $\hat{\Delta}_{AC}^*$ may be unbiased because the $BC$ population is covered within that of $AC$. In MAIC, the excluded subpopulation will receive very low weights (low odds of enrolment in $BC$ vs.~$AC$), while the included subpopulation receives high weights and dominates the reweighted sample. These extreme weights lead to large reductions in ESS and to the deterioration of precision and efficiency. Removing observations from the $AC$ patient-level data, so that inclusion/exclusion criteria are consistent, explicitly lowers the $AC$ sample size and may degrade precision further. Of course, when there is no interpolation or extrapolation overlap whatsoever, MAIC cannot generate a population-adjusted estimate for the treatment effect.  

\subsection*{Correct specification of the $BC$ population}

MAIC and STC make certain assumptions about the joint distribution of covariates in the $BC$ trial. The restriction of limited IPD makes it unlikely that such joint distribution is available. Summary statistics for the marginal distributions are typically published instead. Where no correlation information is available for the $BC$ study, both methods seem to assume that the joint $BC$ covariate distribution is the product of the published marginal distributions. The implicit assumptions are, in fact, more nuanced. 

In MAIC, as stated in the NICE Decision Support Unit Technical Support Document,\cite{phillippo2016nice} ``when covariate correlations are not available from the ($BC$) population, and therefore cannot be balanced by inclusion in the weighting model, they are assumed to be equal to the correlations amongst covariates in the pseudo-population formed by weighting the ($AC$) population.'' In the typical usage of STC (i.e., the ``plug-in'' approach to the method), the assumption differs slightly. The correlations between the $BC$ covariates are assumed to be equal to the correlations between covariates in the $AC$ study. In the ``covariate simulation'' approach to STC, discussed in subsection \ref{subsec32}, this assumption is also made, albeit more explicitly, if the correlation structure observed in the $AC$ IPD is used to simulate the covariates. In an anchored comparison, only effect-modifying covariates need balancing, so the assumptions can be relaxed to only include effect modifiers. This set of assumptions will only induce bias if higher-order interactions (involving two or more covariates) are unaccounted for or misspecified. If these interactions are not included in the weighting model for MAIC or in the outcome regression for STC, the specification of pairwise correlations will not make a difference in terms of bias, as observed in a recent simulation study that investigates this set of assumptions.\cite{phillippo2020multilevel} 

\subsection*{Linear (parametric) modeling assumptions}

Indirect treatment comparisons should be performed and are typically conducted on the linear predictor scale,\cite{phillippo2016nice} upon which the treatment effect is assumed to be additive for all indirect comparisons. In the anchored case, MAIC and STC have assumed that the effect modifiers have been defined on the linear predictor scale and are additive on this scale, but this assumption is not always appropriate. MAIC and STC are subject to scale conflicts or to bias if effect modification status, which is scale-specific, has been justified on the wrong scale, e.g.~when treatment effect modification is specified as linear but is non-linear or multiplicative, e.g.~age in cardiovascular disease treatments.

This form of model misspecification is more evident in a regression adjustment method like STC, where an explicit outcome regression is formulated. The parametric model depends on functional form assumptions that will be violated if the relationship between the covariates and the outcome is not captured correctly. Even though the logistic regression model for the weights in MAIC does not make reference to the outcome, the method is also susceptible to model misspecification bias, albeit in a more implicit form. The model for estimating the weights is approximately correct in the simulation study because the right subset of covariates has been selected as effect modifiers and the balancing property holds for the weights, as mentioned in subsection \ref{subsec44}. In practice, the model will be incorrectly specified if this is not the case, potentially leading to a biased estimate. Note that, in practice, we find that it may be more difficult to specify a correct parametric model for the outcome than an approximately correct parametric model for the trial assignment weights.  

\subsection*{Shared effect modification}

MAIC and STC only produce an estimate $\hat{\Delta}_{AB}^*$ that is valid in the $BC$ population, implicitly assumed to be the target population. This may not match the target population for the decision, unless an additional assumption is made: the shared effect modifier assumption.\cite{phillippo2016nice} Otherwise, one must assume that the target population of the comparison is the $BC$ population.

This assumption is untestable with the available data and implies that: (1) active treatments $A$ and $B$ have the same set of treatment effect modifiers; and (2) the interaction effects of each effect modifier are identical for both treatments. Then, estimate $\hat{\Delta}_{AB}^*$ can be transported to any given target population as effect modifiers are guaranteed to cancel out (relative effects for $A$ vs.~$B$ are conditionally constant across all populations). 

Shared effect modification is hard to meet in practice if the competing interventions do not belong to the same class, and have dissimilar mechanisms of action or clinical properties. In that case, there is little reason to believe that treatments $A$ and $B$ have the same set of effect-modifying covariates and that these interact with active treatment in the same way in $AC$ and $BC$. 

\subsection*{Concluding remarks}

In practice, some of the assumptions above may be hard to meet. Population-adjusted analyses require a larger number of assumptions than standard comparisons. These may be untestable with the available data, and will create additional complexity. Violations of the assumptions may lead to biased estimates of the treatment effect. Hence, it is important that future simulation studies assess the robustness of the methods to failures of the assumptions considered in this article, and under different degrees of data availability and model misspecification.

\clearpage

\section*{Supplementary Appendix B: Simulation study scenario settings}\label{SB}
\addcontentsline{toc}{section}{Supplementary Appendix B: Simulation study scenario settings}

In Table \ref{tab1}, parameter values for each simulation scenario are presented.

\begingroup\fontsize{7}{9}\selectfont
\begin{longtabu} to \linewidth {|>{}r|>{\raggedleft}X>{\raggedleft}X>{\raggedleft}X>{\raggedleft}X>{}r|}
\caption{\label{tab1}Parameter values for the simulation study scenarios.}\\
\hline
Scenario & Number of subjects in \textit{AC} & Prognostic effect & Interaction effect & Mean of \textit{AC} covariates & Covariate correlation\\
\hline
\endfirsthead
\caption[]{Parameter values for the simulation study scenarios. \textit{(continued)}}\\
\hline
Scenario & Number of subjects in \textit{AC} & Prognostic effect & Interaction effect & Mean of \textit{AC} covariates & Covariate correlation\\
\hline
\endhead
\rowcolor{gray!6}  1 & 150 & 0.40 & 0.40 & 0.45 & 0.00\\
\hline
2 & 300 & 0.40 & 0.40 & 0.45 & 0.00\\
\hline
\rowcolor{gray!6}  3 & 600 & 0.40 & 0.40 & 0.45 & 0.00\\
\hline
4 & 150 & 0.69 & 0.40 & 0.45 & 0.00\\
\hline
\rowcolor{gray!6}  5 & 300 & 0.69 & 0.40 & 0.45 & 0.00\\
\hline
6 & 600 & 0.69 & 0.40 & 0.45 & 0.00\\
\hline
\rowcolor{gray!6}  7 & 150 & 1.11 & 0.40 & 0.45 & 0.00\\
\hline
8 & 300 & 1.11 & 0.40 & 0.45 & 0.00\\
\hline
\rowcolor{gray!6}  9 & 600 & 1.11 & 0.40 & 0.45 & 0.00\\
\hline
10 & 150 & 0.40 & 0.69 & 0.45 & 0.00\\
\hline
\rowcolor{gray!6}  11 & 300 & 0.40 & 0.69 & 0.45 & 0.00\\
\hline
12 & 600 & 0.40 & 0.69 & 0.45 & 0.00\\
\hline
\rowcolor{gray!6}  13 & 150 & 0.69 & 0.69 & 0.45 & 0.00\\
\hline
14 & 300 & 0.69 & 0.69 & 0.45 & 0.00\\
\hline
\rowcolor{gray!6}  15 & 600 & 0.69 & 0.69 & 0.45 & 0.00\\
\hline
16 & 150 & 1.11 & 0.69 & 0.45 & 0.00\\
\hline
\rowcolor{gray!6}  17 & 300 & 1.11 & 0.69 & 0.45 & 0.00\\
\hline
18 & 600 & 1.11 & 0.69 & 0.45 & 0.00\\
\hline
\rowcolor{gray!6}  19 & 150 & 0.40 & 1.11 & 0.45 & 0.00\\
\hline
20 & 300 & 0.40 & 1.11 & 0.45 & 0.00\\
\hline
\rowcolor{gray!6}  21 & 600 & 0.40 & 1.11 & 0.45 & 0.00\\
\hline
22 & 150 & 0.69 & 1.11 & 0.45 & 0.00\\
\hline
\rowcolor{gray!6}  23 & 300 & 0.69 & 1.11 & 0.45 & 0.00\\
\hline
24 & 600 & 0.69 & 1.11 & 0.45 & 0.00\\
\hline
\rowcolor{gray!6}  25 & 150 & 1.11 & 1.11 & 0.45 & 0.00\\
\hline
26 & 300 & 1.11 & 1.11 & 0.45 & 0.00\\
\hline
\rowcolor{gray!6}  27 & 600 & 1.11 & 1.11 & 0.45 & 0.00\\
\hline
28 & 150 & 0.40 & 0.40 & 0.45 & 0.35\\
\hline
\rowcolor{gray!6}  29 & 300 & 0.40 & 0.40 & 0.45 & 0.35\\
\hline
30 & 600 & 0.40 & 0.40 & 0.45 & 0.35\\
\hline
\rowcolor{gray!6}  31 & 150 & 0.69 & 0.40 & 0.45 & 0.35\\
\hline
32 & 300 & 0.69 & 0.40 & 0.45 & 0.35\\
\hline
\rowcolor{gray!6}  33 & 600 & 0.69 & 0.40 & 0.45 & 0.35\\
\hline
34 & 150 & 1.11 & 0.40 & 0.45 & 0.35\\
\hline
\rowcolor{gray!6}  35 & 300 & 1.11 & 0.40 & 0.45 & 0.35\\
\hline
36 & 600 & 1.11 & 0.40 & 0.45 & 0.35\\
\hline
\rowcolor{gray!6}  37 & 150 & 0.40 & 0.69 & 0.45 & 0.35\\
\hline
38 & 300 & 0.40 & 0.69 & 0.45 & 0.35\\
\hline
\rowcolor{gray!6}  39 & 600 & 0.40 & 0.69 & 0.45 & 0.35\\
\hline
40 & 150 & 0.69 & 0.69 & 0.45 & 0.35\\
\hline
\rowcolor{gray!6}  41 & 300 & 0.69 & 0.69 & 0.45 & 0.35\\
\hline
42 & 600 & 0.69 & 0.69 & 0.45 & 0.35\\
\hline
\rowcolor{gray!6}  43 & 150 & 1.11 & 0.69 & 0.45 & 0.35\\
\hline
44 & 300 & 1.11 & 0.69 & 0.45 & 0.35\\
\hline
\rowcolor{gray!6}  45 & 600 & 1.11 & 0.69 & 0.45 & 0.35\\
\hline
46 & 150 & 0.40 & 1.11 & 0.45 & 0.35\\
\hline
\rowcolor{gray!6}  47 & 300 & 0.40 & 1.11 & 0.45 & 0.35\\
\hline
48 & 600 & 0.40 & 1.11 & 0.45 & 0.35\\
\hline
\rowcolor{gray!6}  49 & 150 & 0.69 & 1.11 & 0.45 & 0.35\\
\hline
50 & 300 & 0.69 & 1.11 & 0.45 & 0.35\\
\hline
\rowcolor{gray!6}  51 & 600 & 0.69 & 1.11 & 0.45 & 0.35\\
\hline
52 & 150 & 1.11 & 1.11 & 0.45 & 0.35\\
\hline
\rowcolor{gray!6}  53 & 300 & 1.11 & 1.11 & 0.45 & 0.35\\
\hline
54 & 600 & 1.11 & 1.11 & 0.45 & 0.35\\
\hline
\rowcolor{gray!6}  55 & 150 & 0.40 & 0.40 & 0.30 & 0.00\\
\hline
56 & 300 & 0.40 & 0.40 & 0.30 & 0.00\\
\hline
\rowcolor{gray!6}  57 & 600 & 0.40 & 0.40 & 0.30 & 0.00\\
\hline
58 & 150 & 0.69 & 0.40 & 0.30 & 0.00\\
\hline
\rowcolor{gray!6}  59 & 300 & 0.69 & 0.40 & 0.30 & 0.00\\
\hline
60 & 600 & 0.69 & 0.40 & 0.30 & 0.00\\
\hline
\rowcolor{gray!6}  61 & 150 & 1.11 & 0.40 & 0.30 & 0.00\\
\hline
62 & 300 & 1.11 & 0.40 & 0.30 & 0.00\\
\hline
\rowcolor{gray!6}  63 & 600 & 1.11 & 0.40 & 0.30 & 0.00\\
\hline
64 & 150 & 0.40 & 0.69 & 0.30 & 0.00\\
\hline
\rowcolor{gray!6}  65 & 300 & 0.40 & 0.69 & 0.30 & 0.00\\
\hline
66 & 600 & 0.40 & 0.69 & 0.30 & 0.00\\
\hline
\rowcolor{gray!6}  67 & 150 & 0.69 & 0.69 & 0.30 & 0.00\\
\hline
68 & 300 & 0.69 & 0.69 & 0.30 & 0.00\\
\hline
\rowcolor{gray!6}  69 & 600 & 0.69 & 0.69 & 0.30 & 0.00\\
\hline
70 & 150 & 1.11 & 0.69 & 0.30 & 0.00\\
\hline
\rowcolor{gray!6}  71 & 300 & 1.11 & 0.69 & 0.30 & 0.00\\
\hline
72 & 600 & 1.11 & 0.69 & 0.30 & 0.00\\
\hline
\rowcolor{gray!6}  73 & 150 & 0.40 & 1.11 & 0.30 & 0.00\\
\hline
74 & 300 & 0.40 & 1.11 & 0.30 & 0.00\\
\hline
\rowcolor{gray!6}  75 & 600 & 0.40 & 1.11 & 0.30 & 0.00\\
\hline
76 & 150 & 0.69 & 1.11 & 0.30 & 0.00\\
\hline
\rowcolor{gray!6}  77 & 300 & 0.69 & 1.11 & 0.30 & 0.00\\
\hline
78 & 600 & 0.69 & 1.11 & 0.30 & 0.00\\
\hline
\rowcolor{gray!6}  79 & 150 & 1.11 & 1.11 & 0.30 & 0.00\\
\hline
80 & 300 & 1.11 & 1.11 & 0.30 & 0.00\\
\hline
\rowcolor{gray!6}  81 & 600 & 1.11 & 1.11 & 0.30 & 0.00\\
\hline
82 & 150 & 0.40 & 0.40 & 0.30 & 0.35\\
\hline
\rowcolor{gray!6}  83 & 300 & 0.40 & 0.40 & 0.30 & 0.35\\
\hline
84 & 600 & 0.40 & 0.40 & 0.30 & 0.35\\
\hline
\rowcolor{gray!6}  85 & 150 & 0.69 & 0.40 & 0.30 & 0.35\\
\hline
86 & 300 & 0.69 & 0.40 & 0.30 & 0.35\\
\hline
\rowcolor{gray!6}  87 & 600 & 0.69 & 0.40 & 0.30 & 0.35\\
\hline
88 & 150 & 1.11 & 0.40 & 0.30 & 0.35\\
\hline
\rowcolor{gray!6}  89 & 300 & 1.11 & 0.40 & 0.30 & 0.35\\
\hline
90 & 600 & 1.11 & 0.40 & 0.30 & 0.35\\
\hline
\rowcolor{gray!6}  91 & 150 & 0.40 & 0.69 & 0.30 & 0.35\\
\hline
92 & 300 & 0.40 & 0.69 & 0.30 & 0.35\\
\hline
\rowcolor{gray!6}  93 & 600 & 0.40 & 0.69 & 0.30 & 0.35\\
\hline
94 & 150 & 0.69 & 0.69 & 0.30 & 0.35\\
\hline
\rowcolor{gray!6}  95 & 300 & 0.69 & 0.69 & 0.30 & 0.35\\
\hline
96 & 600 & 0.69 & 0.69 & 0.30 & 0.35\\
\hline
\rowcolor{gray!6}  97 & 150 & 1.11 & 0.69 & 0.30 & 0.35\\
\hline
98 & 300 & 1.11 & 0.69 & 0.30 & 0.35\\
\hline
\rowcolor{gray!6}  99 & 600 & 1.11 & 0.69 & 0.30 & 0.35\\
\hline
100 & 150 & 0.40 & 1.11 & 0.30 & 0.35\\
\hline
\rowcolor{gray!6}  101 & 300 & 0.40 & 1.11 & 0.30 & 0.35\\
\hline
102 & 600 & 0.40 & 1.11 & 0.30 & 0.35\\
\hline
\rowcolor{gray!6}  103 & 150 & 0.69 & 1.11 & 0.30 & 0.35\\
\hline
104 & 300 & 0.69 & 1.11 & 0.30 & 0.35\\
\hline
\rowcolor{gray!6}  105 & 600 & 0.69 & 1.11 & 0.30 & 0.35\\
\hline
106 & 150 & 1.11 & 1.11 & 0.30 & 0.35\\
\hline
\rowcolor{gray!6}  107 & 300 & 1.11 & 1.11 & 0.30 & 0.35\\
\hline
108 & 600 & 1.11 & 1.11 & 0.30 & 0.35\\
\hline
\rowcolor{gray!6}  109 & 150 & 0.40 & 0.40 & 0.15 & 0.00\\
\hline
110 & 300 & 0.40 & 0.40 & 0.15 & 0.00\\
\hline
\rowcolor{gray!6}  111 & 600 & 0.40 & 0.40 & 0.15 & 0.00\\
\hline
112 & 150 & 0.69 & 0.40 & 0.15 & 0.00\\
\hline
\rowcolor{gray!6}  113 & 300 & 0.69 & 0.40 & 0.15 & 0.00\\
\hline
114 & 600 & 0.69 & 0.40 & 0.15 & 0.00\\
\hline
\rowcolor{gray!6}  115 & 150 & 1.11 & 0.40 & 0.15 & 0.00\\
\hline
116 & 300 & 1.11 & 0.40 & 0.15 & 0.00\\
\hline
\rowcolor{gray!6}  117 & 600 & 1.11 & 0.40 & 0.15 & 0.00\\
\hline
118 & 150 & 0.40 & 0.69 & 0.15 & 0.00\\
\hline
\rowcolor{gray!6}  119 & 300 & 0.40 & 0.69 & 0.15 & 0.00\\
\hline
120 & 600 & 0.40 & 0.69 & 0.15 & 0.00\\
\hline
\rowcolor{gray!6}  121 & 150 & 0.69 & 0.69 & 0.15 & 0.00\\
\hline
122 & 300 & 0.69 & 0.69 & 0.15 & 0.00\\
\hline
\rowcolor{gray!6}  123 & 600 & 0.69 & 0.69 & 0.15 & 0.00\\
\hline
124 & 150 & 1.11 & 0.69 & 0.15 & 0.00\\
\hline
\rowcolor{gray!6}  125 & 300 & 1.11 & 0.69 & 0.15 & 0.00\\
\hline
126 & 600 & 1.11 & 0.69 & 0.15 & 0.00\\
\hline
\rowcolor{gray!6}  127 & 150 & 0.40 & 1.11 & 0.15 & 0.00\\
\hline
128 & 300 & 0.40 & 1.11 & 0.15 & 0.00\\
\hline
\rowcolor{gray!6}  129 & 600 & 0.40 & 1.11 & 0.15 & 0.00\\
\hline
130 & 150 & 0.69 & 1.11 & 0.15 & 0.00\\
\hline
\rowcolor{gray!6}  131 & 300 & 0.69 & 1.11 & 0.15 & 0.00\\
\hline
132 & 600 & 0.69 & 1.11 & 0.15 & 0.00\\
\hline
\rowcolor{gray!6}  133 & 150 & 1.11 & 1.11 & 0.15 & 0.00\\
\hline
134 & 300 & 1.11 & 1.11 & 0.15 & 0.00\\
\hline
\rowcolor{gray!6}  135 & 600 & 1.11 & 1.11 & 0.15 & 0.00\\
\hline
136 & 150 & 0.40 & 0.40 & 0.15 & 0.35\\
\hline
\rowcolor{gray!6}  137 & 300 & 0.40 & 0.40 & 0.15 & 0.35\\
\hline
138 & 600 & 0.40 & 0.40 & 0.15 & 0.35\\
\hline
\rowcolor{gray!6}  139 & 150 & 0.69 & 0.40 & 0.15 & 0.35\\
\hline
140 & 300 & 0.69 & 0.40 & 0.15 & 0.35\\
\hline
\rowcolor{gray!6}  141 & 600 & 0.69 & 0.40 & 0.15 & 0.35\\
\hline
142 & 150 & 1.11 & 0.40 & 0.15 & 0.35\\
\hline
\rowcolor{gray!6}  143 & 300 & 1.11 & 0.40 & 0.15 & 0.35\\
\hline
144 & 600 & 1.11 & 0.40 & 0.15 & 0.35\\
\hline
\rowcolor{gray!6}  145 & 150 & 0.40 & 0.69 & 0.15 & 0.35\\
\hline
146 & 300 & 0.40 & 0.69 & 0.15 & 0.35\\
\hline
\rowcolor{gray!6}  147 & 600 & 0.40 & 0.69 & 0.15 & 0.35\\
\hline
148 & 150 & 0.69 & 0.69 & 0.15 & 0.35\\
\hline
\rowcolor{gray!6}  149 & 300 & 0.69 & 0.69 & 0.15 & 0.35\\
\hline
150 & 600 & 0.69 & 0.69 & 0.15 & 0.35\\
\hline
\rowcolor{gray!6}  151 & 150 & 1.11 & 0.69 & 0.15 & 0.35\\
\hline
152 & 300 & 1.11 & 0.69 & 0.15 & 0.35\\
\hline
\rowcolor{gray!6}  153 & 600 & 1.11 & 0.69 & 0.15 & 0.35\\
\hline
154 & 150 & 0.40 & 1.11 & 0.15 & 0.35\\
\hline
\rowcolor{gray!6}  155 & 300 & 0.40 & 1.11 & 0.15 & 0.35\\
\hline
156 & 600 & 0.40 & 1.11 & 0.15 & 0.35\\
\hline
\rowcolor{gray!6}  157 & 150 & 0.69 & 1.11 & 0.15 & 0.35\\
\hline
158 & 300 & 0.69 & 1.11 & 0.15 & 0.35\\
\hline
\rowcolor{gray!6}  159 & 600 & 0.69 & 1.11 & 0.15 & 0.35\\
\hline
160 & 150 & 1.11 & 1.11 & 0.15 & 0.35\\
\hline
\rowcolor{gray!6}  161 & 300 & 1.11 & 1.11 & 0.15 & 0.35\\
\hline
162 & 600 & 1.11 & 1.11 & 0.15 & 0.35\\
\hline
\end{longtabu}
\endgroup{}

\clearpage

\section*{Supplementary Appendix C: Simulation study results}\label{SC}
\addcontentsline{toc}{section}{Supplementary Appendix C: Simulation study results}

Table \ref{tab2} displays the key performance measures associated with each of the simulated scenarios. Biases are displayed in \textcolor{red}{red} when their absolute size is greater than one half of the treatment effect estimate's empirical standard error. Coverage rates are presented in \textcolor{red}{red} when these are statistically significantly different to 0.95; namely, if the rate is less than 0.9365 or more than 0.9635. Monte Carlo standard errors for each measure are presented in parentheses.

\setlength{\LTleft}{-20cm plus -1fill}
\setlength{\LTright}{\LTleft}
\begingroup\fontsize{6}{8}\selectfont
\begin{longtable}[t]{llllllllll}
\caption{\label{tab2}Performance metrics for each method and simulation scenario. Monte Carlo standard errors for each measure are presented in parentheses. ATE: average estimated marginal treatment effect for $A$ vs.~$B$ (is equal to the bias as the true effect is zero); LCI: average lower bound of the 95 percent confidence interval; UCI: average upper bound of the 95 percent confidence interval; MSE: mean square error; MAE: mean absolute error; Cover: coverage rate of the 95 percent confidence intervals; VR: variability ratio; ESE: empirical standard error; MAIC: matching-adjusted indirect comparison; STC: simulated treatment comparison.}\\
\toprule
Scenario & Method & ATE & LCI & UCI & MSE & MAE & Cover & VR & ESE\\
\midrule
\endfirsthead
\caption[]{Performance metrics for each method and simulation scenario. Monte Carlo standard errors for each measure are presented in parentheses. ATE: average estimated marginal treatment effect for $A$ vs.~$B$ (is equal to the bias as the true effect is zero); LCI: average lower bound of the 95 percent confidence interval; UCI: average upper bound of the 95 percent confidence interval; MSE: mean square error; MAE: mean absolute error; Cover: coverage rate of the 95 percent confidence intervals; VR: variability ratio; ESE: empirical standard error; MAIC: matching-adjusted indirect comparison; STC: simulated treatment comparison. \textit{(continued)}}\\
\toprule
Scenario & Method & ATE & LCI & UCI & MSE & MAE & Cover & VR & ESE\\
\midrule
\endhead
\
\endfoot
\bottomrule
\endlastfoot
\rowcolor{gray!6}  1 & MAIC & \textcolor{black}{-0.010 (0.007)} & -0.429 (0.007) & 0.410 (0.007) & 0.046 (0.002) & 0.170 (0.007) & \textcolor{black}{0.950 (0.007)} & 1.001 (0.022) & 0.214 (0.005)\\
1 & STC & \textcolor{black}{-0.065 (0.007)} & -0.513 (0.008) & 0.383 (0.007) & 0.059 (0.003) & 0.195 (0.007) & \textcolor{black}{0.938 (0.008)} & 0.974 (0.022) & 0.235 (0.005)\\
\rowcolor{gray!6}  1 & Bucher & \textcolor{black}{-0.095 (0.007)} & -0.504 (0.007) & 0.315 (0.006) & 0.054 (0.002) & 0.187 (0.007) & \textcolor{red}{0.927 (0.008)} & 0.985 (0.022) & 0.212 (0.005)\\
\addlinespace
2 & MAIC & \textcolor{black}{0.010 (0.005)} & -0.310 (0.005) & 0.331 (0.005) & 0.026 (0.001) & 0.129 (0.005) & \textcolor{black}{0.955 (0.007)} & 1.020 (0.023) & 0.160 (0.004)\\
\rowcolor{gray!6}  2 & STC & \textcolor{black}{-0.034 (0.005)} & -0.369 (0.005) & 0.300 (0.005) & 0.030 (0.001) & 0.140 (0.005) & \textcolor{black}{0.950 (0.007)} & 1.006 (0.023) & 0.170 (0.004)\\
2 & Bucher & \textcolor{black}{-0.075 (0.005)} & -0.390 (0.005) & 0.239 (0.005) & 0.031 (0.001) & 0.141 (0.005) & \textcolor{red}{0.928 (0.008)} & 1.007 (0.023) & 0.159 (0.004)\\
\addlinespace
\rowcolor{gray!6}  3 & MAIC & \textcolor{black}{-0.009 (0.004)} & -0.267 (0.004) & 0.250 (0.004) & 0.018 (0.001) & 0.108 (0.004) & \textcolor{black}{0.939 (0.008)} & 0.975 (0.022) & 0.135 (0.003)\\
3 & STC & \textcolor{black}{-0.046 (0.004)} & -0.312 (0.004) & 0.220 (0.004) & 0.022 (0.001) & 0.120 (0.004) & \textcolor{red}{0.924 (0.008)} & 0.964 (0.022) & 0.141 (0.003)\\
\end{longtable}
\endgroup{}

\pagebreak

\section*{Supplementary Appendix D: Example code}\label{SD}
\addcontentsline{toc}{section}{Supplementary Appendix D: Example code}

Example \texttt{R} code implementing MAIC, the conventional version of STC and the Bucher method on a simulated example is provided in this appendix. The code and data are available at \url{https://github.com/remiroazocar/population_adjustment_simstudy} in the \texttt{Example} subdirectory. Full code for the simulation study is available in the online repository. The simulation study and the provided example use survival outcomes, with a Cox proportional hazards regression as the outcome model of interest in the analysis.

\subsection*{MAIC}

\begin{lstlisting}
library("survival") # required for weighted Cox regression

AC.IPD <- read.csv("Example/AC_IPD.csv") # load AC patient-level data
BC.ALD <- read.csv("Example/BC_ALD.csv") # load BC aggregate-level data

N <- nrow(AC.IPD) # number of subjects in AC
X.EM <- AC.IPD[,c("X1","X2")] # AC effect modifiers 
bar.X.EM.BC <- BC.ALD[,c("mean_X1", "mean_X2")] # BC effect modifier means
K.EM <- ncol(X.EM) # number of effect modifiers 

# center the AC effect modifiers on the BC means
for (k in 1:K.EM) {
  X.EM[,k] <- X.EM[,k] - bar.X.EM.BC[,k]
}

# objective function to be minimized for weight estimation
Q <- function(alpha, X.EM) {
  return(sum(exp(X.EM %*% alpha)))
}

alpha <- rep(1,K.EM) # arbitrary starting point for the optimiser
# objective function minimized using BFGS
Q.min <- optim(fn=Q, X.EM=as.matrix(X.EM), par=alpha, method="BFGS")
hat.alpha <- Q.min$par # finite solution is the logistic regression parameters
log.hat.w <- rep(0, N)
for (k in 1:K.EM) {
  log.hat.w <- log.hat.w + hat.alpha[k]*X.EM[,k]
}
hat.w <- exp(log.hat.w) # estimated weights 
aess <- sum(hat.w)^2/sum(hat.w^2) # approximate effective sample size

# fit weighted Cox proportional hazards model using robust=TRUE for robust variance
outcome.fit <- coxph(Surv(time, status)~trt, robust=TRUE, weights=hat.w,data=AC.IPD)

# fitted treatment coefficient is relative effect for A vs. C
hat.Delta.AC <- summary(outcome.fit)$coef[1] 
hat.var.Delta.AC <- vcov(outcome.fit)[[1]] # estimated variance for A vs. C
hat.Delta.BC <- with(BC.ALD, logHR_B) # B vs. C
hat.var.Delta.BC <- with(BC.ALD, var_logHR_B)
hat.Delta.AB <- hat.Delta.AC - hat.Delta.BC # A vs. B
hat.var.Delta.AB <- hat.var.Delta.AC + hat.var.Delta.BC
# construct Wald-type normal distribution-based confidence interval
uci.Delta.AB <- hat.Delta.AB + qnorm(0.975)*sqrt(hat.var.Delta.AB)
lci.Delta.AB <- hat.Delta.AB + qnorm(0.025)*sqrt(hat.var.Delta.AB) 
\end{lstlisting}

\subsection*{Conventional STC}

\begin{lstlisting}
library("survival") # required for standard Cox regression

AC.IPD <- read.csv("Example/AC_IPD.csv") # load AC patient-level data
BC.ALD <- read.csv("Example/BC_ALD.csv") # load BC aggregate-level data

# fit regression of outcome on the baseline characteristics and treatment
# effect modifiers are centered at the mean BC values
# purely prognostic variables are included but not centered
outcome.fit <- coxph(Surv(time, status)~X3+X4+trt*I(X1-BC.ALD$mean_X1)+trt*I(X2-BC.ALD$mean_X2),
                     data=AC.IPD)

# estimated treatment coefficient is relative effect for A vs. C
hat.Delta.AC <- coef(outcome.fit)["trt"]
hat.var.Delta.AC <- vcov(outcome.fit)["trt", "trt"] # estimated variance for A vs. C
hat.Delta.BC <- with(BC.ALD, logHR_B) # B vs. C
hat.var.Delta.BC <- with(BC.ALD, var_logHR_B)
hat.Delta.AB <- hat.Delta.AC - hat.Delta.BC # A vs. B
hat.var.Delta.AB <- hat.var.Delta.AC + hat.var.Delta.BC
# construct Wald-type normal distribution-based confidence interval
uci.Delta.AB <- hat.Delta.AB + qnorm(0.975)*sqrt(hat.var.Delta.AB)
lci.Delta.AB <- hat.Delta.AB + qnorm(0.025)*sqrt(hat.var.Delta.AB) 
\end{lstlisting}

\subsection*{Bucher method}

\begin{lstlisting}
library("survival") # required for standard Cox regression

AC.IPD <- read.csv("Example/AC_IPD.csv") # load AC patient-level data
BC.ALD <- read.csv("Example/BC_ALD.csv") # load BC aggregate-level data

# simple regression of outcome on treatment
outcome.fit <- coxph(Surv(time, status)~trt, data=AC.IPD)

# fitted treatment coefficient is relative effect for A vs. C
hat.Delta.AC <- coef(outcome.fit)["trt"]
hat.var.Delta.AC <- vcov(outcome.fit)["trt", "trt"] # estimated variance for A vs. C
hat.Delta.BC <- with(BC.ALD, logHR_B) # B vs. C
hat.var.Delta.BC <- with(BC.ALD, var_logHR_B)  
hat.Delta.AB <- hat.Delta.AC - hat.Delta.BC # A vs. B
hat.var.Delta.AB <- hat.var.Delta.AC + hat.var.Delta.BC
# construct Wald-type normal distribution-based confidence interval
uci.Delta.AB <- hat.Delta.AB + qnorm(0.975)*sqrt(hat.var.Delta.AB)
lci.Delta.AB <- hat.Delta.AB + qnorm(0.025)*sqrt(hat.var.Delta.AB) 
\end{lstlisting}


\pagebreak

\bibliographystyle{unsrt}
\addcontentsline{toc}{section}{References}
\bibliography{references}


\end{document}